\documentclass[showpacs,twocolumn,amsmath,amssymb,pra,superscriptaddress]{revtex4}

\usepackage{graphicx}
\usepackage{amsmath,amsfonts,amssymb,amsthm}
\usepackage{fancyhdr}
\usepackage{mathrsfs}
\usepackage{epstopdf}
\usepackage{setspace}
\usepackage{bm}
\usepackage[colorlinks]{hyperref}

\begin{document}

\title{Multiplying and detecting propagating microwave photons using inelastic Cooper-pair tunneling}

\author{Juha Lepp\"akangas}
\affiliation{Institut f\"ur Theoretische Festk\"orperphysik, Karlsruhe Institute of Technology, D-76128 Karlsruhe, Germany}

\author{Michael Marthaler}
\affiliation{Institut f\"ur Theoretische Festk\"orperphysik, Karlsruhe Institute of Technology, D-76128 Karlsruhe, Germany}

\author{Dibyendu Hazra}
\affiliation{Univ.\ Grenoble Alpes, CEA, INAC-Pheliqs, F-38000 Grenoble, France}

\author{Salha Jebari}
\affiliation{Univ.\ Grenoble Alpes, CEA, INAC-Pheliqs, F-38000 Grenoble, France}

\author{Romain Albert}
\affiliation{Univ.\ Grenoble Alpes, CEA, INAC-Pheliqs, F-38000 Grenoble, France}

\author{Florian Blanchet}
\affiliation{Univ.\ Grenoble Alpes, CEA, INAC-Pheliqs, F-38000 Grenoble, France}

\author{G\"oran Johansson}
\affiliation{Microtechnology and Nanoscience, MC2, Chalmers University of Technology, SE-412 96 G\"oteborg, Sweden}

\author{Max Hofheinz}
\affiliation{Univ.\ Grenoble Alpes, CEA, INAC-Pheliqs, F-38000 Grenoble, France}




\begin{abstract}

The interaction between propagating microwave fields and Cooper-pair tunneling across a DC voltage-biased Josephson junction can be highly nonlinear. We show theoretically that this nonlinearity can be used to convert an incoming single microwave photon into an outgoing $n$-photon Fock state in a different mode. In this process, the electrostatic energy released in a Cooper-pair tunneling event is transferred to the outgoing Fock state, providing energy gain.
 The created multi-photon Fock state is frequency entangled and highly bunched.
The conversion can be made reflectionless (impedance-matched) so that all incoming photons are converted to $n$-photon states. With realistic parameters multiplication ratios $n > 2$ can be reached. By two consecutive multiplications,
the outgoing Fock-state number can get sufficiently large to accurately discriminate it from vacuum with linear post-amplification and power measurement. 
Therefore, this amplification scheme can be used as single-photon detector without dead time.
\end{abstract}

\maketitle

\section{Introduction}

The ability to control light at the single-photon level is a key ingredient of most quantum systems in the optical and microwave domain.
In the optical domain, single-photon detectors (SPDs) play a central role: they
are the workhorse of most quantum optics experiments and fundamental research tools, such as quantum state tomography \cite{Lvovsky2009}.
Together with the creation of nonclassical states of light they can also be used for
quantum communication~\cite{WallsMilburn, Gisins} and optical quantum computing~\cite{Menicucci2006,Milburn2007,Langford2011}.
In particular, a SPD together with a photon multiplier facilitates nonlinear optical quantum computing~\cite{Langford2011}.

In the microwave domain, a true SPD of itinerant microwaves has not yet been realized despite important recent developments \cite{Chen2011,Govia2012,Solano2009,Sathyamoorthy2014,Koshino2015,Inomata2016,Kyrienko2016,Royer2017}. Instead, readout of quantum devices 
relies on linear parametric amplifiers \cite{Yurke1988,Castellanos-Beltran2007,Bergeal2010,Macklin2015} with noise levels very close to the standard quantum limit of 1\,photon (including zero-point fluctuations of the incoming line). Unfortunately, this unavoidable noise does not allow them to discriminate between a vacuum state and a single photon propagating along a transmission line (TL). A microwave SPD could do just this and would allow for a host of new possibilities for readout of quantum devices and communication using quantum microwaves.

In this article, we propose building a microwave photon multiplier and
SPD based on the nonlinear coupling between charge tunneling and electromagnetic fields in a microwave circuit. From early on it has been established how this coupling modifies charge transport \cite{Ingold1992,Devoret1990, Girvin1990, Holst1994, Pertti2006}, but recent technological progress now also allows for the measurement of the emitted radiation~\cite{Hofheinz2011,Reulet3,Saira2016,Parlavecchio2015,Cassidy2017,Fabien2017}.
This in turn has stimulated further theoretical studies of its properties~\cite{Leppakangas2013,Leppakangas2016,Quassemi2015,Marthaler2011,Gramich2013,Paris2015,Souquet2016,Leppakangas2015,Dambach2015,Armour2013,Koppenhofer2017,Hassler2015}.
A DC voltage-biased Josephson junction, embedded in a superconducting microwave circuit,
exhibits the strong nonlinearity of this light-charge interaction most clearly, due to the absence of quasi-particle excitations. This system is understood to be a
bright and robust on-chip source of nonclassical microwave radiation, such as of
antibunched photons~\cite{Leppakangas2015,Dambach2015}, nonclassical photon pairs~\cite{Hofheinz2011,Leppakangas2013,Paris2015,Fabien2017}, and multi-photon Fock states~\cite{Souquet2016,Koppenhofer2017}.

\begin{figure}[tb] 
\includegraphics[width=0.8\linewidth]{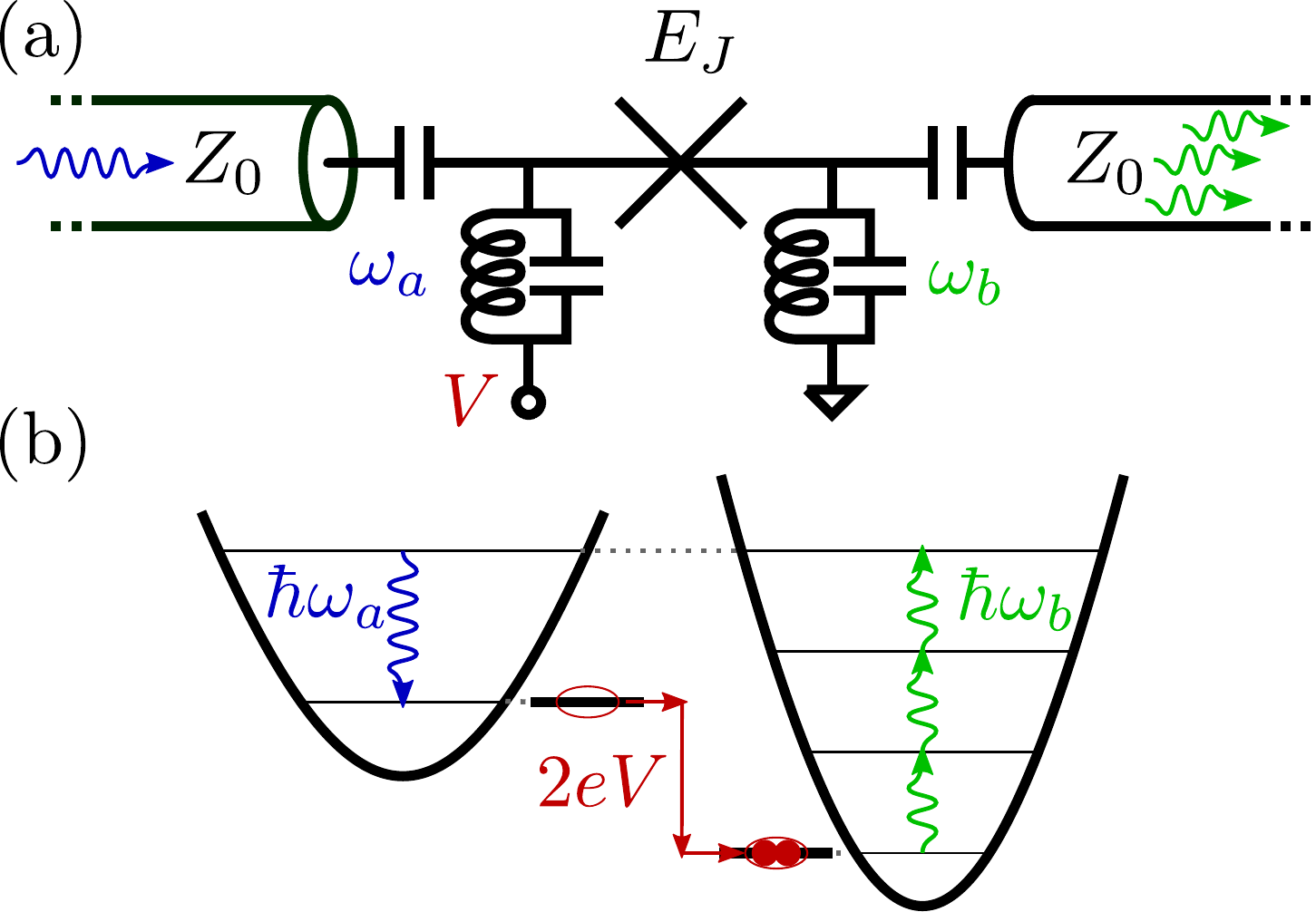}
\caption{(a): 
We investigate microwave scattering in a transmission line connected to two resonators, with
frequencies $\omega_a$ and $\omega_b$, and a Josephson junction with coupling energy $E_{\rm J}$.
An incoming photon from the left interacts with Cooper pair tunneling across the Josephson junction that emits an outgoing field to right.
When impedance matched, an incoming photon of frequency $\omega_{a}$ deterministically converts into $n$ outgoing photons of average frequency $\omega_{b}$.
(b): The energy diagram of photon tripling with slight frequency down-conversion.
Energy is absorbed from the Cooper-pair tunneling event, $\hbar\omega_{a}+ 2eV =3\times\hbar\omega_{b}$.
Generally, it is possible to up-convert ($\omega_{a}>\omega_{b}$) and down-convert ($\omega_{a}<\omega_{b}$) incoming microwave photons.
}
\label{fig:figure1}
\end{figure}

We explore theoretically a process which converts a propagating photon in one mode to $n$ photons in another mode. 
Such nonlinear interaction can be realized in a microwave circuit depicted in Fig.~\ref{fig:figure1}:
a voltage-biased Josephson junction couples two TLs via two microwave resonators at different frequencies. 
Incoming photons from the left-hand side TL interact with the Josephson junction,
which creates a reflected field to the left and a converted field to the right of the Josephson junction.
We show that there exists an impedance-matched situation, where an incoming photon is deterministically absorbed and converted into an outgoing multi-photon Fock state on the right-hand side. 
The energy released in the simultaneous Cooper-pair tunneling event, $2eV$, is absorbed by the creation of $n$ photons,
$2eV+\hbar\omega_a=n\hbar\omega_b$, and thereby allows for energy gain.
The created multi-photon Fock state is frequency entangled and the photon distribution is highly bunched.
Unlike the down-conversion process in a parametric amplifier, this conversion process requires an incoming photon and ideally cannot be triggered by zero-point fluctuations~\cite{Bergeal2010}.
The bias condition is different from other recently studied Josephson systems, producing
microwave lasing~\cite{Cassidy2017} or Casimir radiation~\cite{DCE,Lahteenmaki2013}
through two-photon down-conversion processes triggered by vacuum fluctuations.
Our system therefore offers a new tool to manipulate and convert propagating microwave photons
in microwave circuits, without adding photon noise.

If arbitrary system parameters can be realized, multiplication by any $n$ is possible. However, for
presently achievable characteristic impedances of microwave resonators
$n=3$ photon production from a single-photon input is feasible.
More photons can be created when the process is cascaded by connecting the output of the first multiplier to the input of the second one,
in particular, in an integrated setup with two Josephson junctions and three microwave resonators.
By analyzing quadrature fluctuations of Fock states, we find that two
such multiplications can create enough ($3\times 3=9$) photons to be discriminated from vacuum using linear parametric amplifiers, with quantum efficiency $0.9$ and dark-count rate $10^{-3}\times$~bandwidth.
In comparison to other recent proposals, such as single-photon absorption in a phase-qubit type system \cite{Chen2011,Govia2012,Solano2009},
in a lambda-type system \cite{Koshino2015,Inomata2016}, a driven three-level system \cite{Sathyamoorthy2014,Kyrienko2016}, or using transitions to dark states in multi-qubit system \cite{Royer2017}, our microwave SPD does not include artificial atoms, which need to be reset after each detection.
Our system therefore allows for detection of photons without any dark time. 

The article is organized as follows. In Sec.~\ref{sec:Theory},
we introduce the continuous-mode treatment of the propagating radiation in TLs and boundary conditions
describing their interaction with the two resonators and the Josephson junction.
In Sec.~\ref{sec:SolutionSinglePhoton}, we derive an analytical expression for the single-to-multiphoton scattering matrix.
We use this to derive the conditions for the conversion to be deterministic (reflectionless) and 
study photon bunching and nonclassical frequency correlations of the created out field.
We also show how to linearize and straightforwardly obtain exact results for the conversion probability in general biasing conditions.
In Sec.~\ref{sec:MultiPhotonInput}, we explore amplification of multi-photon inputs and finite-bandwidth wavepackets
by considering incoming coherent-state pulses and applying a master-equation approach.
In Sec.~\ref{sec:ThreeCavitySetup},
we consider a two-stage cascasion scheme that includes two Josephson junctions and three microwave resonators.
We show when deterministic cascaded multiplication of incoming single-photon states is possible.
In Sec.~\ref{sec:Detection}, we discuss how created multi-photon Fock states can be experimentally detected using linear amplifiers and
power measurement.
In Sec.~\ref{sec:Feasibility}, we give estimates for parasitic effects possibly degrading the performance of the SPD,
originating in finite temperature and
spontaneous photon emission (photon noise). 
Conclusions and discussion are given in Sec.~\ref{sec:Conclusions}.

\section{The model}\label{sec:Theory}
In this section, we introduce the continuous-mode treatment of the electromagnetic radiation in the semi-infinite TLs.
We state the boundary conditions describing the interaction between the propagating fields and the two microwave resonators in the narrow-bandwidth approximation and introduce
the Heisenberg equation of motion accounting for resonator-resonator coupling provided by the DC voltage-biased Josephson junction.
A more detailed derivation of these equations is given in Appendix~A.

\subsection{Transmission line operators}\label{sec:TLOperators}
Our starting point is the quantized representation of a propagating electromagnetic field in a superconducting TL~\cite{WallquistPRB,WallsMilburn,Loudon}.
A solution for the magnetic flux field in the left-hand side transmission line can be written as
\begin{eqnarray}\label{PropagatingField1}
&&\hat\Phi(x<0,t)=\sqrt{\frac{\hbar Z_{0}}{4\pi}}\int_0^\infty\frac{d\omega}{\sqrt{\omega}} \times   \\
&&\left[ \hat a_{\rm  in}(\omega)e^{i(k_\omega x-\omega t)}+\hat a_{\rm out}(\omega)e^{i(-k_\omega x-\omega t)}+{\rm H.c.} \right]\, . \nonumber
\end{eqnarray}
Here $x=0$ corresponds to the position of the Josephson junction and the two resonators.
The characteristic impedance $Z_0=\sqrt{L'/C'}$ and wave number $k_\omega=\omega\sqrt{L'C'}$ are defined by the capacitance $C'$ and inductance $L'$ per unit length.
The operator $\hat a_{\rm in(out)}^{\dagger}(\omega)$ creates and the operator $\hat a_{\rm in(out)}(\omega)$ annihilates an
incoming (outgoing) propagating photon of frequency $\omega$.
We have the commutation relations
\begin{equation}\label{CommutationRelation}
\left[ \hat a_{\rm in}(\omega),\hat a_{\rm in}^{\dagger}(\omega') \right]=\delta(\omega-\omega') \, ,
\end{equation}
and similarly for the out-operators.

For the right-hand side transmission line we write similarly ($x>0$)
\begin{eqnarray}\label{PropagatingField2}
&&\hat\Phi(x>0,t)=\sqrt{\frac{\hbar Z_{0}}{4\pi}}\int_0^\infty\frac{d\omega}{\sqrt{\omega}} \times \\
&&\left[ \hat b_{\rm  in}(\omega)e^{i(-k_\omega x-\omega t)}+\hat b_{\rm out}(\omega)e^{i(k_\omega x-\omega t)}+{\rm H.c.} \right]\, , \nonumber
\end{eqnarray}
with analogous relations for the field operators,
\begin{equation}\label{CommutationRelation2}
\left[ \hat b_{\rm in}(\omega),\hat b_{\rm in}^{\dagger}(\omega') \right]=\delta(\omega-\omega') \, .
\end{equation}
The relation between in- and out-operators at the two sides is fixed by the boundary conditions and interaction at the resonators, described
in Sec.~\ref{sec:Boundary}.

In this article we consider situations where frequencies only close to resonance frequencies are relevant. We can then approximate the factor $1/\sqrt{\omega}$
in Eqs.~(\ref{PropagatingField1}) and~(\ref{PropagatingField2}) by the corresponding resonance frequencies~\cite{Loudon}.
For example, for the left-hand side transmission line we then write
\begin{eqnarray}
&&\hat \Phi(x<0,t)=\sqrt{\frac{\hbar Z_{0}}{4\pi \omega_a}}\int_{-\infty}^\infty d\omega \times \\
&& \left[ \hat a_{\rm  in}(\omega)e^{i(k_\omega x-\omega t)}+\hat a_{\rm out}(\omega)e^{i(-k_\omega x-\omega t)}+{\rm H.c.} \right]\, , \nonumber
\end{eqnarray}
and similarly for the right-hand side with factor $1/\sqrt{\omega_{b}}$.
Here, we have also formally extended the lower bound of the integration to $-\infty$, which
can be done when frequencies well below $\omega_a$ have negligible contribution.
Within this approximation we then write
\begin{eqnarray}
\hat \Phi(x<0,t)=\sqrt{\frac{\hbar Z_{0}}{2\omega_{a}}}\left[ \hat a_{\rm  in} (t-x/c)+ \hat a_{\rm out} (t+x/c)+{\rm H.c.} \right],
\end{eqnarray}
where we have defined
\begin{eqnarray}\label{eq:FieldDefinition1}
\hat a_{\rm in/out}(t)=\frac{1}{\sqrt{2\pi}}\int_{-\infty}^\infty d\omega e^{-i\omega t} \hat a_{\rm in/out}(\omega) \, ,
\end{eqnarray}
and $c=1/\sqrt{L'C'}$. We have then
\begin{eqnarray}
[ \hat a_{\rm in}(t),\hat a_{\rm in}^{\dagger}(t') ]=\delta(t-t') \, ,
\end{eqnarray}
and similarly for the out-field operators. 
The inverse transformation has the form
\begin{eqnarray}\label{eq:FieldDefinition2}
\hat a_{\rm in/out}(\omega)=\frac{1}{\sqrt{2\pi}}\int_{-\infty}^\infty dt e^{i\omega t} \hat a_{\rm in/out}(t).
\end{eqnarray}
The operator $\hat a_{\rm in}^{(\dagger)}(t)$ annihilates  (creates) an incoming photon at $x=0$ at time $t$.
Analogue definition is made for the  right-hand side transmission line operators $\hat \Phi(x>0,t)$  and $\hat b_{\rm in/out}(t)$.

\subsection{Boundary conditions and Heisenberg equations of motion}\label{sec:Boundary}
The semi-infinite TLs are connected to two resonators, as shown in Fig.~\ref{fig:figure1}. These impose boundary conditions of the form (Appendix~A)
\begin{eqnarray}\label{eq:CavityBoundaryCondition1}
\hat a_{\rm in}(t)+\hat a_{\rm out}(t)&=&\sqrt{\gamma_a}\hat a(t) \\
\hat b_{\rm in}(t)+ \hat b_{\rm out}(t)&=&\sqrt{\gamma_b}\hat b(t)\label{eq:CavityBoundaryCondition2} \, .
\end{eqnarray}
These are time-dependent operators as the boundary conditions are given in the Heisenberg picture.
The photon annihilation (creation) operator $\hat a^{(\dagger)}$ corresponds to the standard description of the local field in the left-hand side resonator and $\hat b^{(\dagger)}$ in the right-hand side resonator.
We have $[\hat a,\hat a^\dagger]=1$ and $[\hat b,\hat b^\dagger]=1$, other combinations of these operators vanish.
The energy decay rate $\gamma_{a/b}$ of the cavity field in the corresponding TL
defines the bandwidth of the resonator $a/b$ (we assume that there is no internal dissipation of resonators).

The field operators additionally follow the Heisenberg equations of motion (Appendix~A)
\begin{eqnarray}
\dot{\hat a}(t)&=& \frac{i}{\hbar}\left[H_0+H_{\rm J} ,\hat a(t)\right]  - \frac{\gamma_a}{2}\hat a(t)+\sqrt{\gamma_a}\hat a_{\rm in}(t)\\
\dot{\hat b}(t)&=&  \frac{i}{\hbar}\left[H_0+H_{\rm J} ,\hat b(t)\right]  - \frac{\gamma_b}{2}\hat b(t)+\sqrt{\gamma_b}\hat b_{\rm in}(t)\label{eq:Heisenberg2}.
\end{eqnarray}
Here $H_0=\hbar\omega_{a}\hat a^\dagger \hat a + \hbar\omega_{b}\hat b^\dagger \hat b$ is the resonator Hamiltonian.
The interaction between them is provided by the Josephson junction Hamiltonian,
\begin{equation}
H_{\rm J}=-E_{\rm J}\cos\left[\omega_{\rm J}t+g_a(\hat a+\hat a^\dagger)-g_b(\hat b+\hat b^\dagger)  \right],
\end{equation}
where the Josephson frequency $\omega_{\rm J}=2eV/\hbar$ accounts for the DC voltage bias
and the dimensionless coupling $g_{a/b}=\sqrt{\pi Z_{a/b}/R_{\rm Q}}$ compares the characteristic impedances
of modes $a$ and $b$ to the resistance quantum $R_{\rm Q}=h/4e^2$.

In the following sections, the above boundary conditions and equations of motion are used to evaluate certain
expectation values for the out field using specific inputs, under rotating-wave approximation (RWA).
More precisely, in Sec.~\ref{sec:SolutionSinglePhoton}, we show an exact analytical solution for the scattering matrix when having a single-photon input.
In Sec.~\ref{sec:MultiPhotonInput}, we study conversion of multi-photon inputs by considering incoming coherent-state pulses.
In Sec.~\ref{sec:ThreeCavitySetup}, we study double multiplication of single incoming photons with two cascaded multipliers.
Finally, in Sec.~\ref{sec:Feasibility}, we estimate perturbatively the effect of vacuum and thermal fluctuations at other frequencies,
which were neglected when taking the RWA and the narrow-bandwidth approximation.

\section{Single-photon input and deterministic multiplication}\label{sec:SolutionSinglePhoton}
In this section, we consider single-photon input of the photomultiplier. We first evaluate the single-to-multi-photon scattering matrix
and then show how to linearize the problem and derive results for general conversion probabilities and bandwidths.
We also study the quantum information carried by the created propagating multi-photon states.
In particular, the created states are found to exhibit frequency and time-bin entanglement
and carry quantum information of the input state.
We solve the problem for a single-photon input in the rotating wave approximation (RWA). 
Within this model, we treat the cavity and the transmission line exactly and thereby account for the vacuum noise at the resonator frequencies.



\subsection{Scattering matrix (frequency correlations)}\label{sec:ScatteringMatrix}
For a single-photon input at frequency $\omega_{\rm in}\approx\omega_a$ and for a resonant voltage bias $\omega_{\rm J}=n\omega_b-\omega_a$,
we can simplify the Josephson junction Hamiltonian by taking the RWA
(conditions for the validity of this approximation are studied more detailed in Sec.~\ref{sec:Feasibility}). The Hamiltonian becomes
\begin{eqnarray}\label{eq:InteractionHamiltonian}
H^{\rm RWA}_{\rm J}=\hbar\epsilon_{\rm I} \hat a \left(\hat b^\dagger\right)^n e^{-i\omega_{\rm J}t} + {\rm H.c.} \, .
\end{eqnarray}
This creates $n$ photons to oscillator $b$ from a single photon in oscillator $a$, and vice versa.
The amplitude of this process is
\begin{eqnarray}\label{eq:ParametrEpsilon}
\epsilon_{\rm I}= \frac{E_{\rm J}}{2\hbar }\ \frac{i^{n+1}}{n!} \ g_ag_b^{n}e^{-g_a^2/2-g_b^2/2} \, .
\end{eqnarray}

For a single-photon input, we can solve the $n$-photon scattering element analytically.
The Heisenberg equations of motion for the cavity fields have now the form
\begin{eqnarray}
&&\dot{\hat a}(t)= \\
&& -i\omega_a\hat a(t)  - \frac{\gamma_a}{2}\hat a(t)+\sqrt{\gamma_a}\hat a_{\rm in}(t) -i\epsilon_{\rm I}^*\left(\hat b\right)^ne^{+i\omega_{\rm J}t} \nonumber \\
&&\dot{\hat b}(t) = \\
&& -i\omega_b\hat b(t)  - \frac{\gamma_b}{2}\hat b(t)+\sqrt{\gamma_b}\hat b_{\rm in}(t) -in\epsilon_{\rm I}\hat a \left(\hat b^\dagger\right)^{n-1}e^{-i\omega_{\rm J}t} \, . \nonumber
\end{eqnarray}
In the following, we prefer to work with the Fourier-transformed Heisenberg equations of motion.
Using Eqs.~(\ref{eq:FieldDefinition1}) and (\ref{eq:FieldDefinition2}) the Heisenberg equations become then
\begin{widetext}
\begin{eqnarray}\label{eq:Heienberg1}
F_a(\omega)\hat a(\omega)&=& \sqrt{\gamma_a}\hat a_{\rm in}(\omega) -i\frac{\epsilon_{\rm I}^*}{(2\pi)^{(n-1)/2}}\int d\omega_1\ldots\int d\omega_{n-1} \hat b(\omega_1)\ldots \hat b(\omega_{n-1})\hat b(\omega+\omega_{\rm J}-\omega_1 -\ldots -\omega_{n-1})  \\
F_b(\omega)\hat b(\omega)&=& \sqrt{\gamma_b}\hat b_{\rm in}(\omega) -i\frac{n\epsilon_{\rm I}}{(2\pi)^{(n-1)/2}}\int d\omega_1\ldots\int d\omega_{n-1}\hat a(\omega_1)\hat b^\dagger(\omega_2)\ldots b^\dagger(\omega_{n-1})\hat b^\dagger(\omega_1+\omega_{\rm J}-\omega-\omega_2-\ldots-\omega_{n-1})\, , 
\end{eqnarray}
\end{widetext}
where we have defined
\begin{eqnarray}
F_{a/b}(\omega) &=& i(\omega_{a/b}-\omega)+\gamma_{a/b}/2 \, . 
\end{eqnarray}
In these equations, the in-field $\hat a_{\rm in}(\omega)$ [$\hat b_{\rm in}(\omega)$]
can be changed to out-field $-\hat a_{\rm out}(\omega)$ [-$\hat b_{\rm out}(\omega)$]
with simultaneous change $\gamma_{a/b}\rightarrow-\gamma_{a/b}$ in $F_{a/b}(\omega)$.
This is obtained by using the resonator boundary conditions, Eqs.~(\ref{eq:CavityBoundaryCondition1}-\ref{eq:CavityBoundaryCondition2}).

The next step is to determine the scattering matrix
\begin{eqnarray}
A&=&\left\langle 0 \left\vert \hat  b_{\rm out}(\omega_1) \hat b_{\rm out}(\omega_2)\ldots \hat b_{\rm out}(\omega_n)  \hat a^\dagger_{\rm in}(\omega) \right\vert 0 \right\rangle ,
\end{eqnarray}
with the help of resonator boundary conditions and Heisenberg equations of motion.
For simplicity, we will now assume $\omega=\omega_a$ (more general formula is given in Appendix~B).
Using an input-output approach similar to the one developed in Ref.~\cite{Fan2010} we obtain (Appendix~B)
\begin{eqnarray}\label{eq:ResultEntanglement}
&&A=-i\frac{n!}{(2\pi)^{(n-1)/2}} \frac{\epsilon_{\rm I}}{1+\vert\epsilon_n\vert^2}\times \\
&&  \beta(\omega_1)\ldots\beta(\omega_n) \ \alpha(\omega_a) \ \delta(\omega_1+\ldots+\omega_n-\omega_a-\omega_{\rm J})\, . \nonumber
\end{eqnarray}
The dimensionless amplitude $\epsilon_n$ has the form
\begin{eqnarray}
\epsilon_{\rm n}&=&\frac{\epsilon_{\rm I}}{\sqrt{\gamma_a\gamma_b}}\ 2\sqrt{(n-1)!}\, , \label{eq:En}
\end{eqnarray}
and the functions 
\begin{eqnarray}
\alpha(\omega)&=&\frac{\sqrt{\gamma_a}}{i\omega_a-i\omega+\frac{\gamma_a}{2}}=\frac{\sqrt{\gamma_a}}{F_a(\omega)} \\
\beta(\omega)&=&\frac{\sqrt{\gamma_b}}{i\omega_b-i\omega+\frac{\gamma_b}{2}}=\frac{\sqrt{\gamma_b}}{F_b(\omega)} \, ,
\end{eqnarray}
describe the effect of the resonator bandwidths.

\begin{figure}[tb]
\includegraphics[width=\linewidth]{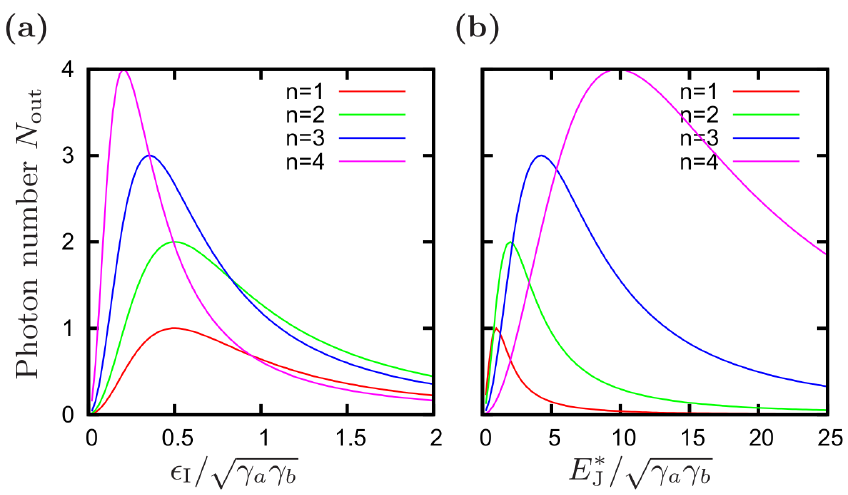}
\caption{
(a): Average number of created photons $N_{\rm out}$ from a single-photon input as a function of (absolute value of) coupling amplitude $\epsilon_{\rm I}$, Eq.~(\ref{eq:Probability}), for multiplication factors $n=1,2,3,4$ (which are also the maximum values of $N_{\rm out}$, correspondingly).
Irrespective of the resonator quality factors, one can always achieve a deterministic photon multiplication (impedance matching) by correctly tuning $\epsilon_{\rm I}$. The corresponding value of $\epsilon_{\rm I}$ decreases with $n$.
(b): The result of (a) plotted as a function of Josephson coupling $E^*_{\rm J}$, Eq.~(\ref{eq:FluxMultiplication}), 
for couplings $g_{a/b}=1$. 
The optimal value for $E_{\rm J}^*$ increases rapidly with~$n$.
This ultimately leads to breakdown of the RWA for higher $n$, as discussed in Sec.~\ref{sec:Feasibility}.
}\label{fig:figure2}
\end{figure}

The average number of outwards propagating photons on side $b$ can also be solved analytically.
We get (assuming incoming photon frequency $\omega_a$)
\begin{eqnarray}\label{eq:Probability}
N_{\rm out}&=&\int d\omega\int d\omega'\left\langle \hat a_{\rm in}(\omega_a) \hat b_{\rm out}^\dagger(\omega') \hat b_{\rm out}(\omega) \hat a^\dagger_{\rm in}(\omega_a)\right\rangle \nonumber\\
&=&n\frac{4 \vert\epsilon_{\rm n}\vert^2}{\left(1+\vert\epsilon_{\rm n}\vert^2 \right)^2}
\end{eqnarray}
We see that when $\vert\epsilon_{\rm n}\vert\rightarrow 0$ or $\vert\epsilon_{\rm n}\vert\rightarrow\infty$,
the incoming field is totally reflected ($N_{\rm out}\rightarrow 0$).
When $\vert\epsilon_{\rm n}\vert=1$, the incoming photon is perfectly converted ($N_{\rm out}=n$).
This reflectionless conversion corresponds to
\begin{equation}\label{eq:FluxMultiplication}
{E^*_{\rm J}}=E_{\rm J}e^{-g_a^2/2-g_b^2/2}=\hbar\sqrt{\gamma_a\gamma_b}\frac{n!}{\sqrt{(n-1)!}g_ag_b^n}\, .
\end{equation}
This central result states that,
irrespective of the resonator quality factors, one can always achieve a deterministic photon multiplication if $E_{\rm J}$ is chosen correctly.
This is visualized in Fig.~\ref{fig:figure2} for the cases $n=1,2,3,4$, both as a function of $\epsilon_{\rm I}$ and $E_{\rm J}^*$
for $g_{a/b}=1$.

The impedance-matching condition of Eq.~(\ref{eq:FluxMultiplication}) is an important result for an experimental realization,
since when the Josephson junction is realized in a SQUID geometry,
the Josephson coupling can be tuned externally to this value via an applied magnetic field.
The practical range of the optimal spot for $E_{\rm J}^*$
is, in realizations considered in this article, of the order of $\hbar\sqrt{\gamma_a\gamma_b}$. 

\subsection{Carried quantum information and the second-order coherence}\label{sec:CarriedQauntumInformation}
The scattering matrix, Eq.~(\ref{eq:ResultEntanglement}), represents a full solution for the single-photon conversion problem (in the RWA) and has interesting non-classical features.
In particular, we find that the created $n$-photon state
is entangled in frequency: It is the superposition of all possible out-field frequency combinations that sum up to $\omega_a+\omega_{\rm J}$,
with amplitudes defined by the cavity broadening factors $\alpha(\omega)$ and $\beta(\omega)$, see Eq.~(\ref{eq:ResultEntanglement}).
This type of correlations are nonclassical and, for example, can violate a Bell inequality for position and time~\cite{Franson1989}.

Furthermore, by Fourier transforming one obtains the shape of the multi-photon Fock state in the time domain \cite{Loudon}.
In case of a two-photon state one gets
\begin{eqnarray}
&&\int d\omega_1\int d\omega_2e^{i(\omega_1t_1+\omega_2t_2)}\beta(\omega_1)\beta(\omega_2) \times \nonumber \\
&& \delta(\omega_a+\omega_{\rm J}-\omega_1-\omega_2)\propto e^{-\gamma_{b}\vert t_1-t_2\vert /2} \, . \nonumber
\end{eqnarray}
For a narrow input bandwidth $\Delta$ but wide $\gamma_b$ the output is therefore highly bunched. 
Further evidence for this is obtained by evaluating the second-order coherence
\begin{eqnarray}
g^{(2)}(\tau)=\frac{G^{(2)}(\tau)}{\vert G^{(1) }(0)\vert^2} \, ,
\end{eqnarray}
where the first-order coherence for propagating fields is here defined as~\cite{Leppakangas2016}
\begin{eqnarray}
G^{(1)}(\tau)= \frac{\hbar Z_0}{4\pi}\int d\omega \int d\omega'\sqrt{\omega\omega'}e^{i\omega \tau}\left\langle \hat b_{\rm out}^\dagger(\omega) \hat b_{\rm out}(\omega') \right\rangle   \, ,
\end{eqnarray}
and the second-order coherence similarly,
\begin{eqnarray}
G^{(2)}(\tau)&=& \left( \frac{\hbar Z_0}{4\pi} \right)^2 \int d\omega \int d\omega'\int d\omega'' \int d\omega'''  \\
&\times &\sqrt{\omega\omega'\omega''\omega'''}  e^{i\tau(\omega'-\omega'')} \nonumber \\
&\times &\left\langle \hat b_{\rm out}^{\dagger}(\omega)\hat  b_{\rm out}^{\dagger}(\omega')\hat  b_{\rm out}(\omega'') \hat b_{\rm out}(\omega''') \right\rangle \, . \nonumber
\end{eqnarray}
We obtain (Appendix C)
\begin{eqnarray}
g^{(2)}(\tau)\propto \left( 1- \frac{1}{n}  \right) \frac{\gamma_b}{\Delta}  e^{-\gamma_b\tau} \, .
\end{eqnarray}
Here $\Delta$ is the (frequency) bandwidth of an incoming single-photon wavepacket (assuming $\gamma_b\gg \Delta$),
converted to the multi-photon Fock state.
This results states that the $n$ photons in the out field appear within time $1/\gamma_b$ from each other,
even though the overall wavepacket is distributed in time as $1/\Delta\gg 1/\gamma_b$. This strong bunching is the precursor of the ``click'' of a single photon detector, which can be seen as a photon multiplier with large gain $n$.

We can also deduce that the superposition of a vacuum and single-photon state, $c_0\vert 0\rangle_{\rm in} +c_1\vert 1\rangle_{\rm in}$,
converts (for $\vert \epsilon_n\vert=1$) into state $c_0\vert 0\rangle_{\rm out} +i^nc_1 \vert n_{\rm entangled}\rangle_{\rm out}$.
This means that the amplification is coherent. Information of the phase of the initial state is transferred to the common phase of the created multi-photon state.
Therefore, quantum information is transferred to the whole ensemble of photons, but not to individual photons.
In a realistic setup, however, the phase of the multi-photon state also suffers from stochastic diffusion
due to low-frequency voltage fluctuations~\cite{Ingold1992} affecting the phase of $\epsilon_n$. Therefore the phase-information will likely be lost in a real device.  In Section~\ref{sec:Feasibility}, we analyze the effect of such voltage fluctuations on the conversion probability.

\subsection{Linearization approach  and input bandwidth}\label{sec:Linearization}
If we are only interested in the probability of multiplication, and not in the exact form of the frequency correlations,
we can solve the problem more straightforwardly with the following linearization approach.
The results derived here agree with the scattering-matrix approach used in Section~\ref{sec:ScatteringMatrix}
(which was also able to capture the exact frequency correlations of the out field).
The linearization on the other hand gives easily access to the input bandwidth.

\subsubsection{Solution for linear conversion ($n=1$)}\label{sec:LinearConversion}
We consider first the case $n=1$ and later map the general solution to this simple case.
After Fourier transformation, the Heisenberg equations of motion become
\begin{eqnarray}
F_a(\omega) \hat a(\omega)&=& \sqrt{\gamma_a}\hat a_{\rm in}(\omega) -i\epsilon_{\rm I}^* \hat b(\omega+\omega_{\rm J}) \\
F_b(\omega)\hat b(\omega)&=& \sqrt{\gamma_b}\hat b_{\rm in}(\omega) -i\epsilon_{\rm I}\hat a(\omega-\omega_{\rm J})\, . 
\end{eqnarray}
The solution satisfies
\begin{eqnarray}
&&\hat b(\omega)\left[ F_b(\omega)+\frac{\vert\epsilon_{\rm I}\vert^2}{i\omega_{\rm J}+F_a(\omega)}  \right] \nonumber \\
&&=\sqrt{\gamma_b}\hat b_{\rm in}(\omega)-i\frac{\epsilon_{\rm I}\sqrt{\gamma_a}\hat a_{\rm in}(\omega-\omega_{\rm J})}{i\omega_{\rm J}+F_a(\omega)}\, . 
\end{eqnarray}
We assume now that there is no input from side $b$. In this case, the outgoing photon flux to side $b$
can be deduced from the relation
\begin{eqnarray}
&&\left\langle \hat b_{\rm out}^\dagger (\omega) \hat b_{\rm out}(\omega') \right\rangle=\gamma_b \left\langle \hat b^\dagger (\omega) \hat b(\omega') \right\rangle \\
&&=\frac{\gamma_a\gamma_b\vert \epsilon_{\rm I}\vert^2}{\left\vert \vert \epsilon_{\rm I}\vert^2+F_a(\omega-\omega_{\rm J})F_b(\omega)  \right\vert^2}\left\langle \hat a^\dagger_{\rm in} (\omega-\omega_{\rm J}) \hat a_{\rm in}(\omega'-\omega_{\rm J}) \right\rangle \, , \nonumber
\end{eqnarray}
where we use the fact that here only $\omega=\omega'$ contributes.
We then obtain the transmission probability for an incoming photon of frequency $\omega$,
\begin{eqnarray}\label{eq:MainResultLinearization}
T&=& \frac{N_{\rm out}}{n}=\frac{1}{n}\frac{ \left\langle \hat b^\dagger_{\rm out}(t) \hat  b_{\rm out}(t) \right\rangle }{ \left\langle \hat a^\dagger_{\rm in}(t) \hat a_{\rm in}(t) \right\rangle }\\
&=&\frac{\gamma_a\gamma_b\vert \epsilon_{\rm I}\vert^2}{\left\vert \vert \epsilon_{\rm I}\vert^2+F_a(\omega)F_b(\omega+\omega_{\rm J})  \right\vert^2}. \nonumber
\end{eqnarray}
When $\omega=\omega_a$, and when the resonance $\omega+\omega_{\rm J}=\omega_b$ is met, we obtain
\begin{equation}\label{eq:Transmission1}
T=\frac{4\vert \epsilon_1\vert^2}{(1+\vert \epsilon_1\vert^2)^2}\, .
\end{equation}
This is the result of Eq.~(\ref{eq:Probability}) (for $n=1$).

Using the general solution, we can now also straightforwardly estimate the effect of voltage-bias offset.
When $\omega=\omega_a$ and $\omega+\omega_{\rm J}=\omega_b+\delta\omega$, describing the effect of bias voltage offset (from the resonance condition),
we obtain
\begin{equation}\label{eq:Transmission1}
T=\frac{4\vert \epsilon_1\vert^2}{(1+\vert \epsilon_1\vert^2)^2+\frac{4\delta\omega^2}{\gamma_b^2}}.
\end{equation}
We see that a voltage offset decreases the conversion probability. In the case $\vert \epsilon_1\vert^2=1$ we
get a Lorentzian form with width defined by the cavity $b$ decay rate, $T=1/(1+\delta\omega^2/\gamma_b^2)$.

The dependence on the input frequency can be deduced by setting an offset $\omega=\omega_a+\delta\omega$,
and keeping the resonance voltage bias condition, leading to $\omega+\omega_{\rm J}=\omega_b+\delta\omega$. This gives us
\begin{eqnarray}\label{eq:ResultBandwidthLeadingOrder}
T=\frac{4\vert\epsilon_1\vert^2}{(1+\vert\epsilon_1\vert^2   -4\frac{\delta\omega^2}{\gamma_a\gamma_b}  )^2+ 4\left(\frac{\delta\omega(\gamma_a+\gamma_b)}{\gamma_a \gamma_b}  \right)^2   }\, .  
\end{eqnarray}
Assuming $\vert\epsilon_1\vert=1$ and  $\gamma_a=\gamma_b$, we get a 4-th order ``rectangular'' filter function
\begin{equation}\label{eq:ResultBandwidthN1Symmetric}
T=\frac{1}{1+4x^4},
\end{equation}
where $x=\delta\omega/\gamma_a$. For $\gamma_a\ll\gamma_b$ we instead get a Lorentzian filter
\begin{equation}\label{eq:ResultBandwidthN1Asymmetric}
T=\frac{1}{1+x^2  }\, . 
\end{equation}

\begin{figure}[tb]
\includegraphics[width=\linewidth]{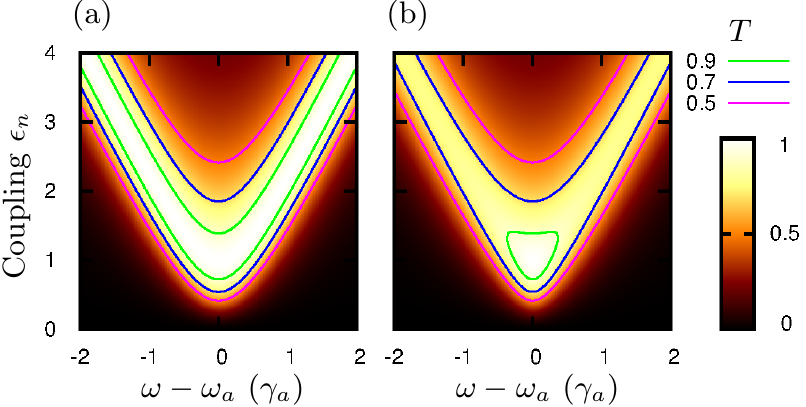}
\caption{
The conversion probability $T$ as a function of frequency offset $\delta\omega=\omega-\omega_a$ and coupling $\epsilon_n$
for  $n=1$ as given by Eq.~(\ref{eq:ResultBandwidthLeadingOrder}).
We consider the cases (a)  $\gamma_a=\gamma_b$ and (b) $\gamma_a=\gamma_b/3$.
We obtain that perfect transmission is also possible for $\vert\epsilon_1\vert> 1$, but only when $\gamma_a=\gamma_b$.
The result of (a) is also valid for arbitrary $n$ when $\gamma_a=n\gamma_b$.
The result of (b) is also valid for $n=3$ when $\gamma_a=\gamma_b$.
}\label{fig:figureSplitting}
\end{figure}

For general $\vert\epsilon_1\vert$, the conversion probability is plotted in Fig.~\ref{fig:figureSplitting}(a) (for $\gamma_a=\gamma_b$).
We observe that when $\vert\epsilon_1\vert>1$, the transmission peak splits into two. 
For $\vert\epsilon_1\vert\gg 1$, we get in good approximation
\begin{eqnarray}\label{eq:ResultBandwidthLeadingOrder2}
  T=\frac{4\gamma_a\gamma_b}{16\left(\delta\omega-\omega_s\right)^2 +  (\gamma_a+\gamma_b)^2}  \, ,  
\end{eqnarray}
where the Lorentzian has the mean
\begin{equation}
  \omega_s=\frac{\vert\epsilon_1\vert\sqrt{\gamma_a\gamma_b}}{2} \,.
\end{equation}
The full-width at half maximum is then $(\gamma_a+\gamma_b)/2$.
Note that perfect transmission for $\vert\epsilon_1\vert> 1$ is possible only when $\gamma_a=\gamma_b$.

\subsubsection{Solution for arbitrary $n$}\label{sec:SolutionWeakInput}
In the case of single-photon input, the preceding results can be straightforwardly generalized to arbitrary $n$, because the resonators can be treated as two-level systems.
Resonator $b$ can be modeled as two level system consisting of 0 and $n$-photon states, because when the photon number drops from $n$ to $n-1$ (due to dissipation in the right-hand side transmission line),
there is no way for the resonator $a$ to be repopulated, and the remaining $n-1$ photons, as well, will inevitably be dissipated in the transmission line $b$.
The effective decay rate of the excited state of the two-level system $b$ is then the one
from the state $n$ to the state $n-1$, that is
\begin{eqnarray}
\tilde\gamma_b=n\gamma_{b} \, .
\end{eqnarray}
Similarly, the effective coupling between the two-level systems is
\begin{eqnarray}
\tilde\epsilon=\epsilon_{I} \sqrt{n!} \, .
\end{eqnarray}
Also the effective resonance frequency can be set to $n\omega_b$, but plays here only the role of a trivial frequency shift.
The final equations of motion are linear and the solution of Eq.~(\ref{eq:MainResultLinearization}) is valid.
The photon multiplication probability is then
\begin{eqnarray}\label{eq:MainResultLinearization2}
T=\frac{\gamma_a\tilde\gamma_b\vert \tilde\epsilon\vert^2}{\left\vert \vert \tilde\epsilon\vert^2+F_a(\omega)\tilde F_b(\omega+\omega_{\rm J})  \right\vert^2},
\end{eqnarray}
where $\tilde F_b(\omega)$ is evaluated using the decay $\tilde\gamma_b=n\gamma_{b}$ and resonance frequency $n\omega_b$.
In the case $\omega=\omega_a$ and resonance condition $\omega+\omega_{\rm J}=n\omega_b$ we get 
\begin{eqnarray}\label{eq:Transmission1}
T&=&\frac{1}{n}\frac{ \left\langle \hat b^\dagger_{\rm out}(t) \hat  b_{\rm out}(t) \right\rangle }{ \left\langle \hat a^\dagger_{\rm in}(t) \hat a_{\rm in}(t) \right\rangle }=\frac{4 \vert\epsilon_n\vert^2}{(1+\vert\epsilon_n\vert^2)^2} \, . \label{eq:Transmission2}
\end{eqnarray}
This is again consistent with the result of Eq.~(\ref{eq:Probability}).

For a general bias voltage offset $\delta\omega$ (see above) we get
\begin{equation}\label{eq:TransmissionGeneral}
T=\frac{4 \vert\epsilon_n\vert^2}{(1+\vert\epsilon_n\vert^2)^2+\frac{4\delta\omega^2}{n^2\gamma_b^2}}.
\end{equation}
In the case $\vert\epsilon_n\vert^2=1$ we have $T=1/(1+\delta\omega^2/n^2\gamma_b^2)$.
Here as well, a voltage offset decreases the conversion probability. 
The probability distribution is again a Lorentzian, but larger with width $n\gamma_b$.


The input bandwidth of the multiplier can again be deduced by using the result of Eq.~(\ref{eq:MainResultLinearization2})
and setting an offset $\omega=\omega_a+\delta\omega$ with $\omega+\omega_{\rm J}=n\omega_b+\delta\omega$. This gives
\begin{eqnarray}\label{eq:ResultBandwidth}
T= \frac{4\vert\epsilon_n\vert^2}{(1+\vert\epsilon_n\vert^2   -4\frac{\delta\omega^2}{\gamma_a\tilde\gamma_b}  )^2+ 4\left(\frac{\delta\omega(\gamma_a+\tilde\gamma_b)}{\gamma_a \tilde\gamma_b}  \right)^2   }\, . 
\end{eqnarray}
Again, the result is just a rescaled function of the case $n=1$, Eq.~(\ref{eq:ResultBandwidthLeadingOrder}).
In particular, if $\gamma_a=n\gamma_b$, we have splitting of the peak as in Fig.~\ref{fig:figureSplitting}(a).
If $\gamma_b=\gamma_a$ and $n=3$, we have splitting as in Fig.~\ref{fig:figureSplitting}(b).
Furthermore, in the case of impedance matching, $\vert\epsilon_n\vert^2=1$ and $\gamma_a=\gamma_b$, we get
\begin{equation}\label{eq:ResultBandwidth2}
T=\frac{1}{1+x^2\left(1-\frac{1}{n}\right)^2 +4\frac{x^4}{n^2}  },
\end{equation}
where we defined $x=\delta\omega/\gamma_a$.
The limiting cases are $T=1/(1+4x^4)$ for $n=1$ and $T=1/(1+x^2)$ for large $n$. The
full width at half maximum (FWHM) changes here from $\sqrt{2}\gamma$ ($n=1$) to $2\gamma$ ($n\gg 1$).

We now summarize the important relations obtained for the widths and forms of the transmission (filter) functions
nearby bias points providing deterministic conversion
\begin{eqnarray}
{\rm FWHM} &=& \gamma_a \,\,\, \,\,\, \,\,\, \,\,\, \,\, (\vert\epsilon_n\vert\gg 1\,\, ,\, \gamma_a=n\gamma_b)\label{eq:FWHM1} \\
{\rm FWHM}&=& \sqrt{2}\gamma_a \,\,\, \,\,\,  (\vert\epsilon_n\vert= 1\,\,\, ,\,\gamma_a=n\gamma_b) \label{eq:FWHM2} \\ 
{\rm FWHM}&=& 2\gamma_a \,\,\, \,\,\, \,\,\, \,\, (\vert\epsilon_n\vert= 1\,\,\, ,\,\gamma_a=\gamma_b \, , \, n\gg 1)\label{eq:FWHM3} \, .
\end{eqnarray}
Note that the filter function [the shape of Eq.~(\ref{eq:ResultBandwidth}) as a function of $\delta\omega$] for the case of Eq.~(\ref{eq:FWHM2})
is more a rectangular than for the two other cases.


\section{Multi-photon input: Coherent state pulses}\label{sec:MultiPhotonInput}
In this section,
we analyze how this multiplier amplifies input signals of higher photon numbers. 
We explore the conversion of coherent-state pulses with varying width and photon number,
and investigate the effect of junction nonlinearities (couplings $g_{a/b}$) and couplings to the transmission lines.

\subsection{Generalized Josephson Hamiltonian}
To account for nonlinear interaction between multi-photon states in  resonators, the Josephson Hamiltonian
(for a resonant bias voltage $\omega_{\rm J} = n\omega_b-\omega_a$) is generalized to~\cite{Wunsche1991}
\begin{eqnarray}\label{eq:GeneralizedJosephsonHamiltonian}
H_{\rm J}^{\rm RWA}&=& (i)^{n+1} \frac{E_{\rm J}}{2}\sum_{k=0}^\infty A_{k+n,k}(g_b)\vert k+n\rangle_b\langle k\vert_b \nonumber  \\
&\times & \sum_{l=0}^\infty A_{l+1,l}(g_a) \vert l\rangle_a\langle l+1\vert_a + {\rm H.c.} \, .
\end{eqnarray}
Here
\begin{eqnarray}
A_{k+n,k}(g)=g^{n} e^{-g^2/2} \sqrt{\frac{k!}{(k+n)!}}L_k^{(n)}(g^2),
\end{eqnarray}
and $L_k^{(n)}(x)$ is the generalized Laguerre polynomial.
Our earlier Hamiltonian, Eq.~(\ref{eq:InteractionHamiltonian}), is obtained
within the approximation $L_k^{(n)}\approx(k+n)!/k!n!$, 
which is exact if $k=0$ (single-photon input). For 
$\sqrt{k} g\gtrsim 1$, the additional nonlinear corrections to the coupling are essential.
In simple terms, unlike for the coupling in Eq.~(\ref{eq:InteractionHamiltonian}), the amplitude does not increase without any limit
when photon numbers increase. The coupling rather oscillates as a function of $\sqrt{k} g\gg 1$~\cite{Marthaler2011,Gramich2013,Souquet2016},
originating in the cosine form of the Josephson energy.
We note that this property is actually beneficial for us, since it allows for better transmission of higher photon-number
inputs.

\subsection{Coherent-state pulses and equivalent master-equation approach}
As the input we consider now specific coherent-state pulses.
We choose a pulse of the form
\begin{eqnarray}\label{eq:CoherentPulse}
\xi(t)= \sqrt{\frac{N_{\rm in}\gamma_{\rm in}}{2}}\exp\left[-i\omega_{a}t -\frac{\gamma_{\rm in} \vert t-t_0\vert}{2} \right]\, . 
\end{eqnarray}
The pulse has on average $\int dt \vert \xi(t)\vert^2 =N_{\rm in}$ photons and at time $t_0$ the peak of the wavepacket reaches the resonator~$a$.
The pulse has a spectral width $\sqrt{\sqrt{2}-1}\gamma_{\rm in} \approx 0.64\gamma_{\rm in}$.


The advantage of coherent-state input is that it allows for a simple master-equation type model for the resonators, because a coherent state input appears as a complex number in the Heisenberg equations for averages.
From these equations we can then deduce the equivalent Lindblad-type master equation.
In this formulation, we have a total Hamiltonian
$\hat H=\hat H_{0}+\hat H_{\rm J}+\hat H_{\rm d}$, where the incoming radiation from side $a$ takes the form of a classical drive,
\begin{eqnarray}
\hat H_{\rm d}=   i\hbar\sqrt{\gamma_{a}}\xi (t) \hat a^\dagger  +  {\rm H.c.} \, . 
\end{eqnarray}
The final equation of motion has the form
\begin{equation}
\hat{\dot {\rho}}=i[\hat \rho,\hat H]+{\cal L}_a[\hat \rho]+{\cal L}_b[\hat \rho] \, ,
\end{equation}
where $\hat \rho$ is the full two-oscillator density matrix and the Lindblad super-operator ${\cal L}_a$
describes decay of field of the oscillator $a$ to the left-hand side transmission line, defined as
\begin{equation}
{\cal L}_a[\hat \rho]=\frac{\gamma_a}{2}\left( 2\hat a \rho \hat a^\dagger -\hat a^\dagger \hat a \rho-\rho\hat a^\dagger \hat a \right)\, .
\end{equation}
Similarly for ${\cal L}_b[\hat \rho]$.

\begin{figure}[tb]
\includegraphics[width=\linewidth]{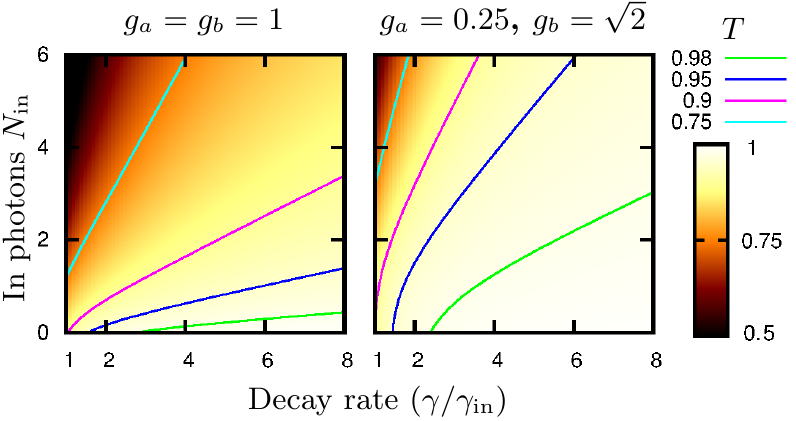}
\caption{
Average conversion probability $T=N_{\rm out}/ nN_{\rm in}$ of coherent state pulses as a function of identical resonator decay rates $\gamma_a=\gamma_b=\gamma$ and average input photon number $N_{\rm in}$, when biased at the photon-tripling resonance ($n=3$).
We consider an incoming pulse of width $\gamma_{\rm in}$ with waveform of Eq.~(\ref{eq:CoherentPulse}).
The reflectionless conversion corresponds to the limit $ \gamma/ \gamma_{\rm in}\rightarrow \infty$, where
$N_{\rm out}/nN_{\rm in}\rightarrow 1$.
}\label{fig:figurepulses}
\end{figure}

\subsection{Numerical results}\label{sec:NumericalResultsTwoCavity}
In Fig.~\ref{fig:figurepulses}(a-b), we plot the numerically evaluated multiplication
efficiency $\left\langle N_{\rm out}\right\rangle /nN_{\rm in}$ as a function of
multiplier bandwidths $\gamma_a=\gamma_b=\gamma$ and the incoming photon number $N_{\rm in}$. We consider 
the case $n=3$, $\vert \epsilon_n \vert =1$ (reflectionless for a single-photon input of frequency $\omega_a$), and
experimentally feasible values (a) $g_a=g_b=1$ and (b) $g_a=0.25$, $g_b=\sqrt{2}$.
We see that the efficiency approaches the ideal value $\left\langle N_{\rm out}\right\rangle /n N_{\rm in}=1$
even for $N_{\rm in}>1$ when $\gamma/\gamma_{\rm in}\rightarrow \infty$, providing deterministic conversion.
In a linear system ($n=1$), the efficiency is a constant for fixed $\gamma/\gamma_{\rm in}$.
However, we see that in the non-linear case ($n>1$) increasing $N_{\rm in}$ decreases the multiplication efficiency.
This means that in the non-linear case ($n>1$) "impedance matching" depends on the photon numbers of the oscillators (and cannot be perfect for a pulse of many photons).
Increase in the decay rate $\gamma$ increases the efficiency, since faster decay keeps the average cavity photon numbers closer to zero.

In Fig.~\ref{fig:figurepulses}(a),
we find roughly a linear dependence between the number of incoming photons $N_{\rm in}$ and bandwidth $\gamma/\gamma_{\rm in}$, when the multiplication efficiency is kept constant (solid contour lines).
For a linear conversion ($n=1$) these lines would be vertical.
We also find that the multiplication efficiency can be increased by decreasing the impedance of the input resonator ($g_a$),
and even more, if we simultaneously increase the output resonator coupling ($g_b$), as shown in Fig.~\ref{fig:figurepulses}(b).
The contour lines are now closer to being vertical, which implies better impedance matching for higher photon-numbers.
The reason is that it is better to keep the input oscillator in the linear regime (small $g_a$) and
instead increase the required nonlinearity of the output resonator (by having $g_b\gtrsim 1$).
The tradeoff for doing this (in comparison to having $g_a=g_b=1$) is a slightly higher rate for emission without input, as shown later in Section~\ref{sec:Feasibility}
(but keeping $g_b=1$ with $g_a=0.25$ would increase the noise essentially more). For $g_{a/b}\ll 1$ (not plotted), we obtain a lower conversion efficiency as shown in Figs.~\ref{fig:figurepulses}(a-b). This regime is also not optimal due to the strong parasitic conversion processes (Section~\ref{sec:Feasibility}).

We conclude that amplification of high photon-number pulses is more efficient when the bandwidth of the multiplier is increased,
which keeps the average photon number in the resonators lower. For experimentally achievable resonator parameters, it is also most efficient when the coupling of the in-resonator $g_{a}\ll 1$ and of the out-resonator $g_{b}\gtrsim 1$. 

\section{Cascaded multiplication: Three-cavity setup}\label{sec:ThreeCavitySetup}
\begin{figure}[tb]
\includegraphics[width=\linewidth]{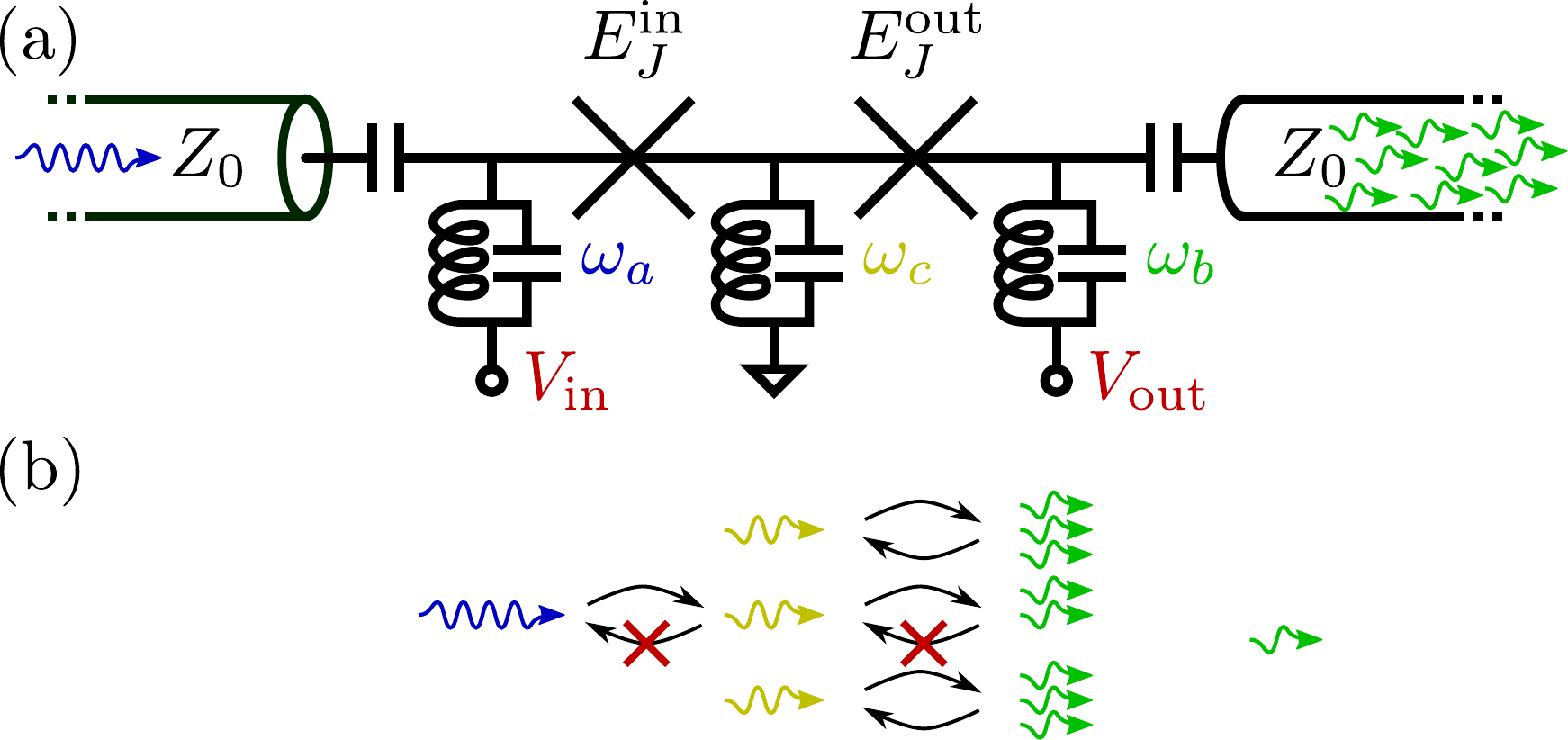}
\caption{Cascaded photomultiplication. (a): Two photomultiplication stages as in Fig.~\ref{fig:figure1} are cascaded with a shared cavity mode at frequency $\omega_c$, acting as output mode for the first stage and as input mode for the second stage. This mode is assumed to have negligible loss. (b): If one photon leaves the output mode, the full process becomes irreversible and all photons have to leave via the output mode. Therefore, like in the one-stage case, an incoming photon is either reflected or fully converted.}\label{fig:scheme2stage}
\end{figure}

In this section we explore two-stage photomultiplication that allows for creating more out photons
from a single-photon input than a single photon multiplier.
This is desired for operation as a single-photon detector, as described in Sec.~\ref{sec:Detection}.
We consider (double) multiplication of incoming single-photon states in a setup
where the output cavity of the first-stage multiplication also acts as the input cavity of the second-state multiplication, see Fig.~\ref{fig:scheme2stage}(a).
One could expect that deterministic photomultiplication becomes more fragile in this more complex setup. On the contrary, we find that deterministic photomultiplication still requires only a single tuning condition. The reason for this constant complexity is that, like in the single junction case, either the incoming photon is reflected or fully converted: As visualized in Fig.~\ref{fig:scheme2stage}(b), as soon as one photon leaves the cavity $b$, 
the resonant backward process (with $n^2$ photon absorption) is no longer possible and
the full process becomes irreversible. In this case all converted photons must leave the system via the output mode.
Therefore, it is sufficient to cancel input reflection via one tuning parameter.

\subsection{Hamiltonian and boundary conditions}
The system we consider includes two Josephson junctions, which we call now the in and the out Josephson junction, separated by a central cavity $c$,
see Fig.~\ref{fig:scheme2stage}(a). 
The Hamiltonian describing this system is a straightforward expansion of the model used in previous sections. We write
\begin{eqnarray}
H=H_{\rm J}^{\rm in}+H_{\rm J}^{\rm out}+H_0 \, ,
\end{eqnarray}
where the Josephson in-Hamiltonian has the form
\begin{equation}
H_{\rm J}^{\rm in}=-E_{\rm J}^{\rm in}\cos\left[\omega^{\rm in}_{\rm J}t+g_a(\hat a+\hat a^\dagger)-g_c(\hat b+\hat b^\dagger)  \right]\, ,
\end{equation}
and the Josephson out-Hamiltonian
\begin{equation}
H_{\rm J}^{\rm out}=-E^{\rm out}_{\rm J}\cos\left[\omega^{\rm out}_{\rm J}t+g_b(\hat b+\hat b^\dagger)-g_c(\hat c+\hat c^\dagger)  \right] \, .
\end{equation}
The two Josephson frequencies account for different voltage biases of the islands, $\hbar\omega_{\rm J}^{\rm in}=2eV_{\rm in}$
and $\hbar\omega_{\rm J}^{\rm out}=2eV_{\rm out}$.
The free evolution resonator Hamiltonian is now
\begin{eqnarray}
H_0=\hbar\omega_{a}\hat a^\dagger \hat a + \hbar\omega_{b}\hat b^\dagger \hat b+ \hbar\omega_{c}\hat c^\dagger \hat c \, .
\end{eqnarray}
To keep the notation similar with the single-junction system,
we have marked $\hat a$ as the in-cavity, $\hat c$ as the middle-cavity, and $\hat b$ as the out-cavity annihilation operator.
The in and out cavities couple to the transmission lines, which is described by the boundary conditions
\begin{eqnarray}\label{eq:CavityBoundaryConditionThreeCavity}
\hat a_{\rm in}(t)+\hat a_{\rm out}(t)&=&\sqrt{\gamma_a}\hat a(t) \\
\hat b_{\rm in}(t)+ \hat b_{\rm out}(t)&=&\sqrt{\gamma_b}\hat b(t)\label{eq:CavityBoundaryCondition3} \, .
\end{eqnarray}
The middle cavity (operator $\hat c$) is assumed to be free of decay.

In the following,
we take the RWA generalized to multiphoton populations, as given by Eq.~(\ref{eq:GeneralizedJosephsonHamiltonian}). We call 
$n_{\rm in}$ the multiplication factor of the in junction and $n_{\rm out}$ of the out junction.

\subsection{Linear solution ($n_{\rm in}=n_{\rm out}=1$)}
The transmission across the three-cavity setup shows important qualitative differences
when compared to the two-cavity setup, which are already present in the linear solution ($n_{\rm in}=n_{\rm out}=1$).
The linear solution for the transmission probability through the device can be derived similarly as presented in Sec.~\ref{sec:Linearization} and has the form
\begin{widetext}
\begin{eqnarray}\label{eq:ThreeCavityLinearSolution}
T=\frac {16 \gamma_a \gamma_b \vert \epsilon_{\rm in}\vert ^2 \vert \epsilon_{\rm out} \vert^2} { 4 \left[-(\gamma_a + \gamma_b) \delta\omega^2 + \gamma_b \vert \epsilon_{a}\vert ^2 + \gamma_a \vert \epsilon_{\rm out}\vert ^2\right]^2 +  \omega^2 \left[\gamma_a \gamma_b + 4 (-\delta\omega^2 + \vert \epsilon_{\rm in}\vert ^2 + \vert \epsilon_{\rm out}\vert ^2)\right]^2 } \, .
\end{eqnarray}
\end{widetext}
Here we have defined the parameters $\epsilon_{\rm in/out}$ similarly as in Eq.~(\ref{eq:ParametrEpsilon}) and
$\delta\omega=\omega-\omega_a$. We assume bias conditions $\omega_a+\omega_{\rm J}^{\rm in}=1\times\omega_c$ and
$\omega_c+\omega_{\rm J}^{\rm out}=1\times\omega_b$.

\begin{figure}[tb]
\includegraphics[width=\linewidth]{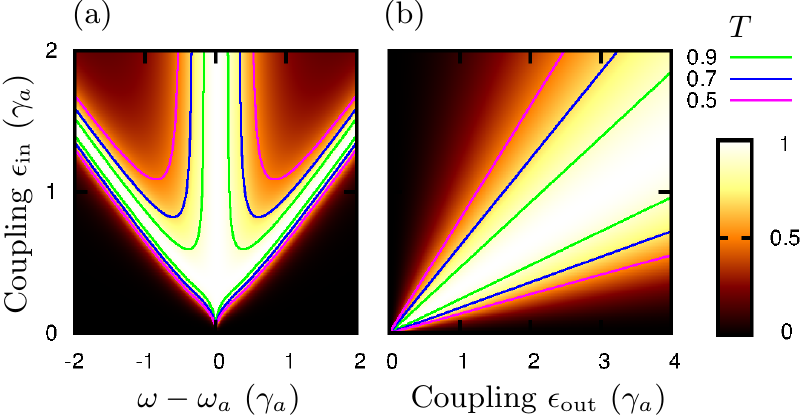} 
\caption{
The conversion probability $T$ in the cascaded setup for $n_{\rm in}=n_{\rm out}=1$, as given by Eq.~(\ref{eq:ThreeCavityLinearSolution}).
(a) For $\gamma_a=\gamma_b$, $\epsilon_{\rm in} = \epsilon_{\rm out}$,
and as a function of frequency offset $\delta\omega=\omega-\omega_a$,
the (deterministic) conversion peak splits into three when the resonator couplings are increased.
(b) For $\omega=\omega_a$ and asymmetric parameters $\gamma_a\neq \gamma_b$ (here $\gamma_b=9\gamma_a$),
deterministic conversion is possible when $\sqrt{\gamma_b}\epsilon_{\rm in} =\sqrt{\gamma_a}  \epsilon_{\rm out} $.
}\label{fig:figureTripling}
\end{figure}

In Fig.~\ref{fig:figureTripling}(a),
the transmission probability is plotted as a function of
$ \epsilon=\vert \epsilon_{\rm in}\vert = \vert \epsilon_{\rm out}\vert $ and input frequency  $\omega$
for $\gamma_a=\gamma_b$.
In Fig.~\ref{fig:figureTripling}(b), the transmission probability is plotted as a function of
 $\vert \epsilon_{\rm in}\vert $ and $ \vert \epsilon_{\rm out}\vert $ when $\gamma_a\ll\gamma_b$ and $\delta\omega=0$.
We see  basically three new features in comparison to the two-cavity setup:
(i) the peak splits into three at $\vert\epsilon\vert\sim\gamma_a/2$ instead of two, (ii) at $\delta\omega=0$ 
perfect transmission is possible for all values of $\vert\epsilon_{\rm in}\vert$ (even when $\gamma_a\neq\gamma_b$),
and (iii) the bandwidth around conversion at $\delta\omega=0$ depends strongly on $\epsilon_{\rm in}$.
We find that all these properties are also present in the nonlinear solution in a very similar form.


The transmission probability at $\delta\omega=0$ can be studied further analytically. Here Eq.~(\ref{eq:ThreeCavityLinearSolution}) gives
\begin{eqnarray}\label{eq:ThreeCavityLinearSolutionZeroOmega}
T=\frac {4 \gamma_a \gamma_b \vert \epsilon_{\rm in}\vert ^2 \vert \epsilon_{\rm out} \vert^2} {  \left( \gamma_b \vert \epsilon_{\rm in}\vert ^2 + \gamma_a \vert \epsilon_{\rm out}\vert ^2\right)^2 } \, .
\end{eqnarray}
We get that deterministic transmission ($T=1$) occurs when
\begin{eqnarray}
 \gamma_b \vert \epsilon_{\rm in}\vert ^2=\gamma_a \vert \epsilon_{\rm out}\vert ^2 \, .
\end{eqnarray}
This means that, for example, increase in the decay of the out-resonator has to be compensated by the increase in the coupling of the out-junction.

The bandwidth for frequencies around $\omega_a$ can also be solved analytically.
We assume now  $\gamma_b \vert \epsilon_{\rm in}\vert ^2=\gamma_a \vert \epsilon_{\rm out}\vert ^2$
and $\gamma_b\gg \gamma_a$. For $\vert\epsilon_{\rm in}\vert<\gamma_a$ and relatively small $\delta\omega$ we get 
\begin{eqnarray}
T\approx\frac{16  \gamma_a^2 \vert \epsilon_{\rm in}\vert^4}{ \gamma_a^4 \delta\omega^2 +  16 \delta\omega^2 \vert \epsilon_{\rm in}\vert^4 + 4 \gamma_a^2 (\delta\omega^4 - 2 \delta\omega^2 \vert \epsilon_{\rm in}\vert^2 + 4 \vert \epsilon_{\rm in}\vert^4)} \, . \nonumber
\end{eqnarray}
We note that for $\vert\epsilon_{\rm in}\vert\ll\gamma_a$ the transmission peak has a Lorentzian form
with width $8\vert \epsilon_{\rm in}\vert^2/\gamma_a$. For  $\vert\epsilon_{\rm in}\vert=\gamma_a/2$ we have a 4-th order ``rectangular'' peak
with width $\sqrt{2}\gamma_a$. For $\vert\epsilon_{\rm in}\vert=\gamma_a$ we have again approximately a Lorentzian form with
width $2\gamma_a$. In summary,
\begin{eqnarray}
{\rm FWHM} &=& \frac{8\vert \epsilon_{\rm in}\vert^2}{\gamma_a} \,\,\, \,\,\, (\vert\epsilon_{\rm in}\vert\ll\gamma_a)\label{eq:FWHM11} \\
{\rm FWHM}&=& \sqrt{2}\gamma_a \,\,\, \,\,\, \,\,\,  (\vert\epsilon_{\rm in}\vert=\frac{\gamma_a}{2}) \label{eq:FWHM12} \\ 
{\rm FWHM}&=& 2\gamma_a \,\,\, \,\,\, \,\,\,  \,\,\, \,\,\, (\vert\epsilon_{\rm in}\vert=\gamma_a)\label{eq:FWHM13} \, .
\end{eqnarray}
Similar relations are found for the case of $n_{\rm in}=n_{\rm out}=3$,
with replacement $\vert\epsilon_{\rm in}\vert/\gamma_a\rightarrow\vert\epsilon_3^{\rm in}\vert$, with the latter variable defined similarly as in Eq.~(\ref{eq:En}).

\subsection{Numerical results for $n_{\rm in}=n_{\rm out}=3$}\label{sec:NumericalResultsThreeCavity}
In the study of conversion probability for cases $n=n_{\rm in}=n_{\rm out}>1$ we resort to numerical methods.
The main feature of the system that helps us solving this problem numerically
is that a single incoming photon needs to be either fully  multiplied by $n^2$,
or fully reflected, see Fig.~\ref{fig:scheme2stage}(b). Other photon numbers in the out field are not allowed.
To obtain the conversion probability, it is then  enough to apply a Lindblad master equation, similar as described in Sec.~\ref{sec:MultiPhotonInput},
using very weak input fields, 
which corresponds to having maximally one photon per time at the input.

\begin{figure}[tb]
\includegraphics[width=\linewidth]{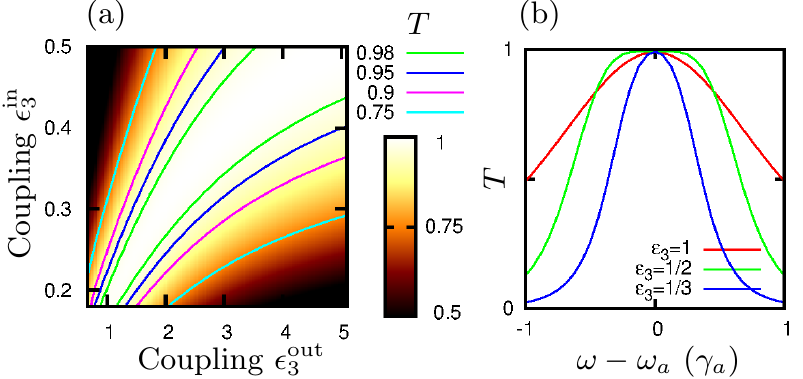} 
\caption{
(a) The conversion probability $T$ in the cascaded photomultiplication when
$n_{\rm in}=n_{\rm out}=3$, $g_c=1.0$, $g_b=1.41$, and $\omega=\omega_a$ (the result does not depend on $g_a$).
The conversion can be deterministic (very close to one) when $\epsilon_{\rm out} \gg \epsilon_{\rm in}$.
(b) The frequency dependence of the transmission probability for $\epsilon^{\rm in}_3=1/3$ (narrow Lorentzian),
$\epsilon^{\rm in}_3=1/2$ (rectangular shape), and $\epsilon^{\rm in}_3=1$ (wide Lorentzian)
with $\epsilon^{\rm out}_3$ that provides deterministic conversion when $g_c=0.25$ and $g_b=1.41$.
}\label{fig:figureThreeCavity}
\end{figure}

In Fig.~\ref{fig:figureThreeCavity}(a), we plot the numerically evaluated single-to-multiphoton conversion probability for the specific case of
$n=3$, converting single incoming photon to 9 outgoing ones. For simplicity, we consider $\gamma_a=\gamma_b$.
We have set a frequency $\omega=\omega_a$ for the incoming field and consider resonant voltage biases $\omega_{\rm J}^{\rm in}+\omega_a=3\times\omega_c$
and $\omega_{\rm J}^{\rm out}+\omega_c=3\times\omega_b$.
We see that similarly as in the linear solution ($n=1$) deterministic multiplication is possible for all values of $\epsilon_{\rm in}$,
if $\epsilon_{\rm out}$ is tuned correctly. In the considered case, $g_c=1.0$ (middle cavity) and $g_b=1.41$ (out cavity),
the out-junction coupling has to be essentially larger than the in-junction coupling: for perfect transmission at $ \vert\epsilon^{\rm in}_{3}\vert =1/2$
(and $\omega=\omega_a$) we need $\vert\epsilon^{\rm out}_3\vert \approx 4$. 
We also find that the out coupling $\vert\epsilon_{\rm out}\vert$ can be reduced by decreasing $g_c$:
for $g_c=0.25$ and $g_b=1.41$ we need approximately $\vert\epsilon_{\rm out}\vert \approx 2\vert\epsilon_{\rm in}\vert $ (not plotted).
The value of $g_a$ does not play a role in this calculation, since the in cavity is populated  maximally by one photon per time.

In Fig.~\ref{fig:figureThreeCavity}(b), we study the conversion bandwidth for three different couplings with $g_c=0.25$ and $g_b=1.41$.
We find that the form of the conversion is very similar to the linear case, Eqs.~(\ref{eq:FWHM11}-\ref{eq:FWHM13}),
within replacement $\vert\epsilon_{\rm in}\vert/\gamma_a\rightarrow\vert\epsilon^{\rm in}_3\vert$, the latter variable as defined in Eq.~(\ref{eq:En}).
In particular, for $\vert\epsilon_3^{\rm in}\vert=1/2$ we obtain a ``rectangular'' shape with width $\approx\sqrt{2}\gamma_a$.
The result is similar also other couplings $g_b$ and $g_c$.

We conclude that also in this system deterministic multiplication can be achieved.
This is possible for all values of Josephson coupling of the first-stage multiplier junction, when the coupling of the second-state
Josephson junction is tuned correctly. The Josephson couplings affect the input bandwidth of the multiplier.

\section{Detection of Fock states using linear amplifiers}\label{sec:Detection}
In this section, we describe how it is possible to transform such a
photomultiplier into a single-photon detector by placing a
quantum-limited phase-preserving amplifier at its output.  The idea
is to measure the instantaneous output power within the output
bandwidth of the photo-multiplier and compare it to a threshold, which
should be high enough to reject the unavoidable noise of the amplifier
due to zero-point fluctuations, but low enough to click when the
photo-multiplier converts an incoming photon to an $n$-photon Fock
state.

\subsection{Power detection of Fock states}

The amplification process produces an $n$-photon Fock state in the output cavity, which then decays into the output mode $b_{\rm out}$ with relaxation rate $\gamma_b$ (see Sec.~\ref{sec:CarriedQauntumInformation}).
We assume this to be true also for the case of cascaded multiplication. 
We now investigate how well such state can be discriminated from vacuum using a quantum-limited phase-preserving linear amplifier.

When a state of the cavity decays into a propagating state in the TL, it gets mixed with vacuum noise of the TL.
As this process is linear, its contribution is known exactly and can be accounted for.
In order to reject as much as possible vacuum noise, the amplifier must then be mode-matched to the output mode of the photo-multiplier (a Lorentzian with width $\gamma_b$). In practice \cite{Eichler2011}, this can be done by choosing an amplifier with a higher bandwidth and numerically convoluting its output with the anti-causal time-domain filter function
\begin{equation}
  f(\tau) = \sqrt{\gamma_b}e^{\gamma_b\tau/2-i\omega_b\tau}\Theta(-\tau)\,,
\end{equation}
where $\Theta(\tau)$ is the Heaviside step function.

Solving Eq.~(\ref{eq:Heisenberg2}) while neglecting the Josephson junction term $H_{\rm J}$ (irreversible decay),
we find how the state of the cavity is related to the vacuum noise of the TL
\begin{eqnarray}
  b(t+t_0) &=& e^{-\gamma_b t/2 -i\omega_bt}b(t_0) \\
  &+& \sqrt{\gamma_b} \int_0^t d\tau e^{-\gamma_b\tau/2} b_{\rm in}(t+t_0-\tau)\,. \nonumber
\end{eqnarray}
The output field, given by Eq.~(\ref{eq:CavityBoundaryCondition2}), convoluted with $f$ is then
\begin{equation}
  \left[b_{\rm out} * f\right](t_0) = b(t_0) \, .
\end{equation}
This means, by mode matching the amplifier to the photo-multiplier output, we can fully reject noise from $b_{\rm in}$ and recover the cavity field $b$ at the input of the amplifier. (However this does not mean that the vacuum noise of the cavity is rejected.)

We now use the fact that the output of a phase-preserving quantum limited amplifier is the scaled Husimi $Q$ function of its input~\cite{Kim97}:
\begin{equation}
  G Q_{{\rm out},t_0}(\sqrt{G} \alpha) = Q_{b(t_0)}(\alpha) \, .
  \label{eq:Qscaling}
  \end{equation}
When the amplifier gain $G$ is large, so that commutators at the output can be neglected, $Q_{\rm out}(\sqrt{G}\alpha)$ directly describes the classical probability density to observe a classical complex amplitude $\sqrt{G}\alpha$ of the amplifier output convoluted with $f$.

The Husimi function $Q_{|n\rangle\langle n|}(\alpha)$ of a n-photon Fock state is independent of the phase of $\alpha$. In order to read the output of the photo-multiplier we therefore calculate the effective photon number $N = |\alpha|^2$ in mode $b$. 
The distribution $D_n(N)$ of measured effective photon number $N$  for a $n$-photon Fock state in mode $b$ is
\begin{equation}
  D_n (N) = \pi Q_{|n\rangle \langle n |} (\sqrt{N}) = \frac{N^n}{n!} e^{-N} \, .
\end{equation}

In order to discriminate between a photon and no-photon, we set a threshold $N_{\rm th}$, with $N < N_{\rm th}$ being interpreted as \lq no click\rq\ and $N \ge N_{\rm th}$ as \lq click\rq.
The probability to get a false click during an inverse bandwidth is:
\begin{equation}
  P_{\rm dark} = \int_{N_{\rm th}}^\infty \operatorname{d}N D_0(N)\, ,
\end{equation}
and the probability to miss a $n$-photon Fock state in mode $b$ is 
\begin{equation}
P_{{\rm miss} |n\rangle}  = \int_0^{N_{\rm th}} \operatorname{d}N D_n(N)\, .
\end{equation}

\begin{figure}  
\includegraphics[width=0.9\columnwidth]{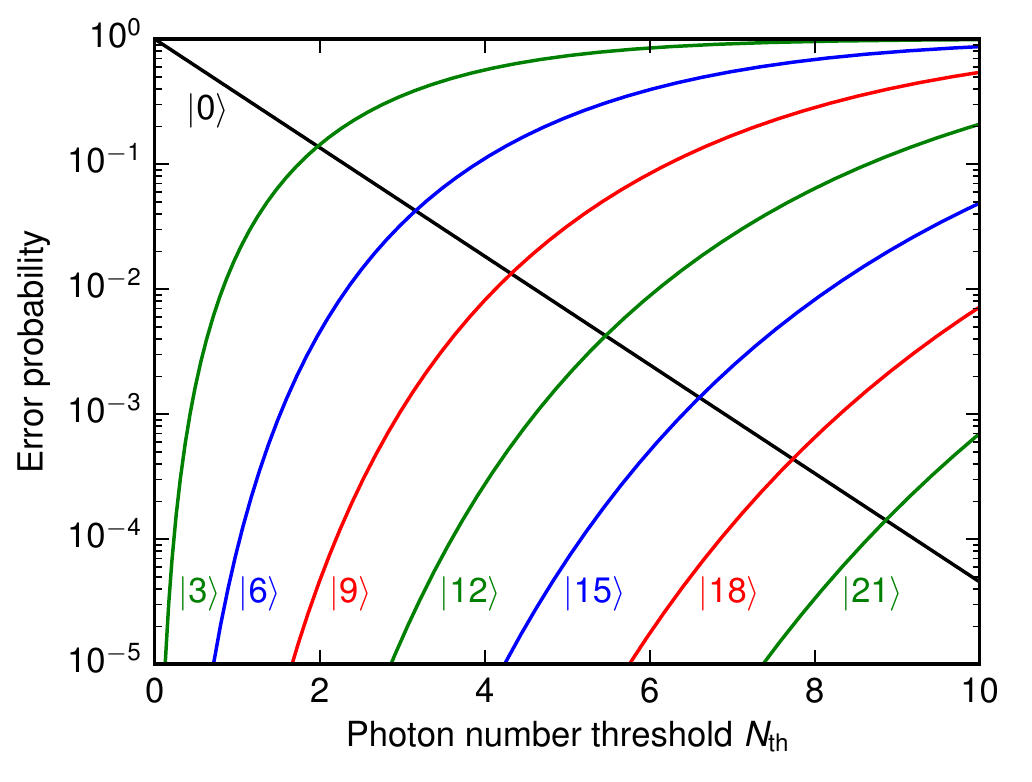}
\caption{Error probabilities for single photon detection using a linear quantum-limited phase-preserving amplifier
at the output of a reflectionless $1 \rightarrow n$ photon multiplier. The detector clicks whenever the effective photon number measured by the amplifier in the output mode of multiplier (see text) exceeds a threshold $N_{\rm th}$. The black line labeled $|0\rangle$ indicates
the dark-count probability $P_{\rm dark}$ (false click) within an inverse bandwidth due to the amplifier noise. The lines labeled $|n\rangle$ with $n > 0$ 
show the probability $P_{{\rm miss} |n\rangle}$ of a $|n\rangle$ state
in the output mode of the photomultiplier not triggering a click.
}\label{fig:darkrate}
\end{figure}

In Fig.~\ref{fig:darkrate}, we plot these error probabilities as a function of the threshold $N_{\rm th}$ and photon multiplication factor $n$. We find that already for a multiplication ratio of $n=3 \times 3=9$ we can obtain a quantum efficiency of approximately 0.9 for a dark count rate of $10^{-3} \times$\,bandwidth.
Much lower dark-count rates for this photon number can be obtained if lower quantum efficiencies are sufficient.
Unlike existing designs~\cite{Inomata2016, Chen2011}, such a SPD can detect another photon immediately after a previous detection event.
We also expect it to be able to resolve photon numbers, even though the efficiency will decrease with photon number, as implied by the numerical results shown in Section~\ref{sec:MultiPhotonInput}.

\section{Feasibility}\label{sec:Feasibility}
So far our analysis has considered an ideal system,
where the RWA and narrow-bandwidth approximation are valid and temperature is zero.
We have also neglected the contribution of Josephson junction capacitance.
In this section, we consider the effect of these contributions for realistic experimental parameters. The thermal and vacuum noise can have two effects: 
Fluctuations at low frequency can bring the device out of the optimal bias condition. Fluctuations at higher frequency can be combined by the nonlinearity of the device to produce emission in the output mode in the absence of input.
Furthermore, a finite junction capacitance provides linear coupling between resonators, which has to be minimized to avoid direct transmission.
Practically, these processes set a lower and higher bound for the input bandwidths of the photo-multiplier.
In the following, we do quantitative noise analysis in the case of single-junction multiplier.
We also use the obtained results to estimate qualitatively the noise in the cascaded (three-cavity) setup.

\subsection{Finite junction capacitance and typical system parameters}\label{sec:JunctionCapacitance}
The junction capacitance $C_{\rm J}$ has been neglected so far in our analysis.
A finite value of $C_{\rm J}$ provides a linear coupling between the resonators, which has to be minimized because it leads to incoming photons being transmitted to the output mode without photon multiplication and frequency conversion.

For the Hamiltonian term describing such (capacitive) coupling we obtain
\begin{eqnarray}
H_{\rm cc}&\approx &\frac{C_{\rm J}}{2\sqrt{(C_{a}+C_{c})(C_{b}+C_{c})}} \hbar \sqrt{\omega_{a}\omega_b}\left( a^\dagger b + ab^\dagger \right) \nonumber \\
&\equiv& g_{cc}\left( a^\dagger b + ab^\dagger \right) \, .
\end{eqnarray}
Here $C_{c}$ is the coupling capacitance between a resonator and a semi-infinite TL and $C_{a/b}$ is the bare resonator capacitance.
The effective resonator $a/b$ capacitance is $C_{a/b}+C_{c}$. We have assumed here $C_{\rm J}\ll C_{a/b}+C_{ c}$.

We can now calculate the probability for direct transmission (without frequency-conversion) through linear coupling by
using the results of Sec.~\ref{sec:LinearConversion}.
Applying Eq.~(\ref{eq:MainResultLinearization}), with identification $\epsilon=g_{cc}$, one obtains
\begin{eqnarray}\label{eq:JunctionCapacitanceTransmission}
T\approx\frac{\gamma_a\gamma_b g_{cc}^2}{g_{cc}^4+\frac{\gamma_a^2}{4} \delta\omega_{\rm r}^2 }\, ,
\end{eqnarray}
where $\delta\omega_{\rm r}=\omega_a-\omega_b$ is the difference between the resonance frequencies
and we have assumed $\vert\delta\omega_{\rm r}\vert\gg\gamma_b$. This function has to be minimized to avoid direct transmission.


For practical parameters of the system, resonator bandwidths $\gamma_{a/b}/2\pi=100$~MHz, couplings $g_{a/b}=1$ (meaning a resonator characteristic impedance $Z_{a/b}=g_{a/b}^2R_{\rm Q}/\pi\approx 2.05$~k$\Omega$), and photon tripling ($n=3$) one obtains
$E_{\rm J}\approx 4.8$~$\mu$eV.
This is an ultrasmall Josephson junction which will have $C_{\rm J}< 1$~fF.
The used parameters give $g_{cc}/h < 200$~MHz.
Then, for example, for $7\rightarrow 5$~GHz conversion  (with photon tripling) we have $\delta\omega_{\rm r}\gg \gamma_{a/b}$ and
Eq.~(\ref{eq:JunctionCapacitanceTransmission}) gives the probability $T < 0.03$ for direct transmission.
Reducing the resonator bandwidths $\gamma_{a/b}$ reduces $T$ further (and also linearly the needed $E_{\rm J}$).
We then conclude that the effect of junction capacitance can be kept negligible.
It can however set an upper limit for the used frequencies,
since higher resonance frequencies demand stronger Josephson couplings (if keeping the quality factors the same),
which then increases the Josephson capacitance. This in turn reduces the resonance frequencies,
since the effective resonator $a/b$ capacitance is $C_{a/b}+C_{\rm J}$.
Due to this tradeoff, we estimate that the scheme is practically extendable up to few tens of GHz (instead of up to the superconducting gap,
which could be over 1~THz). 

\subsection{Effect of thermal fluctuations}
Low-frequency voltage fluctuations are induced by the charge transport as well as finite temperature.
For a low-Ohmic DC bias the effect of temperature is dominating \cite{Leppakangas2014}.
For a bare $50$\,$\Omega$ bias line at $20$\,mK the fluctuations broaden the emission spectrum with the probability distribution
\begin{eqnarray}
P_{\rm lf}(\hbar\delta\omega_{\rm J})\approx \frac{1}{\hbar}\frac{1}{\pi}\frac{\gamma_{\rm thermal}}{\gamma_{\rm thermal}^2+\omega_{\rm J}^2}.
\end{eqnarray}
with $\gamma_{\rm th}/2\pi\approx k_{\rm B} T Z_0 / \hbar R_{\rm Q} \approx 20$\,MHz \cite{Hofheinz2011}. This value can be decreased to $< 4$\,MHz by reducing the value of the impedance at thermally populated frequencies.

To study the effect of such fluctuations, we assume that the voltage fluctuations are adiabatically slow.
We can then use the result for the conversion with bias offset, Eq.~(\ref{eq:TransmissionGeneral}) with $\vert\epsilon_n\vert=1$ 
and $\delta\omega=\omega_{\rm J}-(n\omega_b-\omega_a)$, 
to get the average reflection probability of an incoming photon of frequency $\omega_a$,
\begin{eqnarray}\label{eq:TemperatureReflection}
R &\approx & 1-\int_{-\infty}^\infty d \delta\omega \frac{(n\gamma_b)^2}{(n\gamma_b)^2+\delta\omega^2}P_{\rm lf}(\hbar\delta\omega) \nonumber \\
&=& 1-\frac{n \gamma_b}{\gamma_{\rm thermal}+n\gamma_b}= \frac{\gamma_{\rm thermal}}{\gamma_{\rm thermal}+n\gamma_b}   .
\end{eqnarray}
Therefore, reflection due to low-frequency noise in the voltage is minimized by using resonator bandwidths
and multiplication factors such that $n\gamma_b\gg\gamma_{\rm thermal}$.

\subsection{Spontaneous emission due to vacuum noise} 
The result of Eqs.~(\ref{eq:Probability}-\ref{eq:FluxMultiplication}) implies that any photon number $n$ can be generated from a single-photon input by correctly tuning $\epsilon_{\rm I}$.
However, terms beyond the rotating-wave approximation have been neglected and need to be considered carefully.
For large energy gain the junction must be biased at $2eV=\hbar\omega_{\rm J}>\hbar\omega_b$. Then 
vacuum fluctuations allow for spontaneous emission of one photon to oscillator $b$
and another photon to mode at $\delta\omega =\omega_{\rm J}-\omega_{b}>0$ of the relevant electromagnetic environment
(which was so far neglected)~\cite{Ingold1992,Hofheinz2011}.
It turns out that to keep this effect negligible, 
strong couplings ($g_{a/b}\gtrsim 1$) are needed.
Without specially-engineered high-impedance modes, however, we have $g_i\sim 0.2$ and multi-photon emission is a weak process~\cite{Hofheinz2011,Holst1994,Pertti2006}.
Presently, values $g_i\sim 1$ and slightly beyond can be engineered, for example, by building resonators from high kinetic inductance 
materials\cite{Samkharadze2016} and/or using specific geometries. An alternative approach is to build high-impedance
resonators from Josephson junction arrays~\cite{FabienArray,Stockklauser2017}.


\subsubsection{$P(E)$ approach for estimating the rate of spontaneous emission}

To estimate the emission rate in the output mode without input photons, we can use a perturbative approach in $E_{\rm J}$ developed in Refs.~[\onlinecite{Hofheinz2011,Leppakangas2013,Leppakangas2014}]. 
According to this, the photon flux density (due to thermal and vacuum fluctuations) is of the form
\begin{eqnarray}\label{eq:PEFlux1}
f(\omega)&=&\sum_{\pm}\frac{ 4e^2E_{\rm J}^2 {\rm Re}[Z_{\rm t} (\omega)]}{2\hbar^2\omega}P\left[ \hbar(\pm\omega_{\rm J}-\omega)  \right]\, .   
\end{eqnarray}
Here the well known probability density $P(E)$ is defined as~\cite{Ingold1992}
\begin{equation}
P(E)=\int_{-\infty}^{\infty} dt \frac{1}{2\pi\hbar} e^{J(t)}e^{i\frac{E}{\hbar}t},
\end{equation}
where the phase-correlation function depends on the impedance seen by the tunnel junction, $Z_{\rm t}(\omega)$, as 
\begin{eqnarray}
J(t) &=& \left\langle \left[\hat \phi_0(t)-\hat \phi_0(0)\right]\hat \phi_0(0)\right\rangle\\
\left\langle\hat \phi_0(t)\hat \phi_0(t')\right\rangle&=& 2\int_{-\infty}^{\infty} \frac{d\omega}{\omega}  \frac{ {\rm Re}[Z_{\rm t}(\omega)] }{R_{\rm Q}}\frac{e^{-i\omega(t-t')}}{1-e^{-\beta\hbar\omega}}.
\end{eqnarray}
The two signs in Eq.~(\ref{eq:PEFlux1}) correspond to forward (+) and backward (-) Cooper-pair tunneling events.
As environmental impedance we can consider a Lorentzian resonance at frequency $\omega_b$
\begin{equation}\label{eq:AntiResonance}
{\rm Re}[Z_{\rm t}(\omega)]= \frac{1}{C_b}\frac{\gamma_b}{1+4(\omega-\omega_b)^2\gamma_b^2}\approx \frac{\pi}{2 C_b}\delta(\omega-\omega_b).
\end{equation}
Similarly, we can add this (real part of the) impedance to another Lorentzian peak,
at frequency $\omega_a$, describing resonator $a$.
Finally, we add the resulting function to an assumed background impedance: $50$~Ohm resistor in parallel with capacitance $2$~pF at $T=20$~mK.



\subsubsection{Analytical results}



We first study analytically how to minimize such spontaneous emission.
We use the environmental impedance of Eq.~(\ref{eq:AntiResonance}), which gives
\begin{equation}
P(E)=e^{-g^2_b}\sum_{n=0}^{\infty} \frac{g_b^{2n}}{n!}\delta(E-n\hbar\omega_b).
\end{equation}
We have identified here $g^2_b=(4e^2/2C_b)/\hbar\omega_b$,
assume zero temperature, and consider the limit $\gamma_b\rightarrow 0$~\cite{Ingold1992}.


If we assume that the spontaneous process involves emission of one photon to resonator $b$ and one photon
to frequency $\omega_{\rm J}-\omega_b$, we get that the photon flux at $\omega_b$ (within bandwidth larger than $\gamma_b$)
is proportional to $g^2_b$,
originating in the proportionality to the resonator $b$ impedance in Eq.~(\ref{eq:PEFlux1}). Furthermore,
as the emission rate is proportional to $E_{\rm J}^2$ and the total $P(E)$ to $e^{-g^2_a -g^2_b}$,
the use of Eq.~(\ref{eq:FluxMultiplication}) gives 
\begin{equation}
f_{\omega_b}\equiv\int_{\omega_b} d\omega f(\omega)\propto n\times n!\frac{1}{g_a^2g_b^{2(n-1)}}\, .
\end{equation}
Assuming that at frequency $\omega_{\rm J}-\omega_b$ the impedance contributes with a real number $Z={\rm Re}[Z_{\rm t}(\omega_{\rm J}-\omega_b)]$,
and using the approximation $P(E)=e^{-g^2_a-g^2_b}2Z/R_{\rm Q}\hbar(\omega_{\rm J}-\omega_b)$ in this region, we obtain the photon flux
\begin{eqnarray}\label{eq:Estimate}
f_{\omega_b}=\pi\frac{\gamma_a\gamma_b}{\omega_{\rm J}-\omega_b}\frac{n \times n!}{g_a^2g_b^{2(n-1)}}\frac{Z}{R_{\rm Q}}.
\end{eqnarray}
We find that in order to to reduce spontaneous emission we should always maximize $g_{a/b}$-parameters, particularly the value of $g_b$.
We also see that rates of these spontaneous emission events are also proportional to the real part of the impedance at $\omega_{\rm J}-\omega_b$
(described by the impedance $Z$). This can be reduced by engineering an anti-resonance in the impedance at frequency $\delta\omega=\omega_{\rm J}-\omega_b$.

\subsubsection{Numerical results}\label{sec:NumericalResults}
Eq.~(\ref{eq:Estimate}) is a rough estimate how the decay rate behaves as a function of couplings $g_{a/b}$ and decays $\gamma_{a/b}$.
For more quantitative estimates we need to rely on numerical simulations for specific process and corresponding bias point.
We consider here a bias point providing photon tripling, $7\,\mathrm{GHz} \rightarrow 3\times 5\,\mathrm{GHz}$.

\begin{figure}[tb] 
\includegraphics[width=0.9\linewidth]{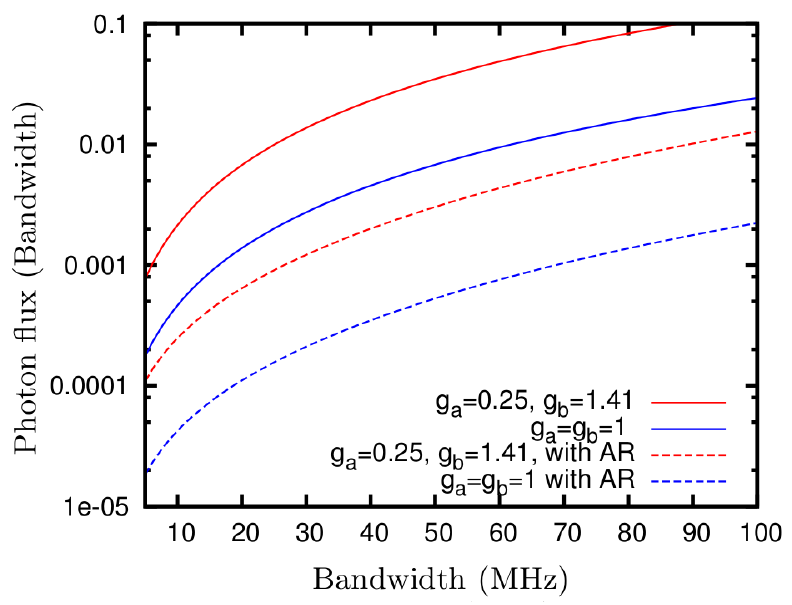}
\caption{ Spontaneous emission by single-junction photomultiplier triggered by vacuum fluctuations at frequencies other than $\omega_{a/b}$.
We consider a bias point for converting single $\omega_a/2\pi=7$~GHz photon into three $\omega_b/2\pi=5$~GHz photons.
We plot the numerically evaluated total photon flux (in units of $\gamma=\gamma_a=\gamma_b$) at output frequency $\omega_b$ within bandwidth 1~GHz
as a function of resonator bandwidths $\gamma/2\pi$. We consider emission for parameters $g_a=0.25$, $g_b=1.41$ (top) and $g_a=g_b=1$ (second from the top).  The emission with an anti-resonance at $\delta\omega = \omega_{\rm J} - \omega_b=3$~GHz
is plotted for $g_a=0.25$, $g_b=1.41$ (third from the top) and $g_a=g_b=1$ (bottom).
The form of the antiresonance is given by Eq.~(\ref{eq:AntiResonance1}).
}\label{fig:spontaneous}
\end{figure}

In Fig.~\ref{fig:spontaneous}, we plot the numerically evaluated spontaneous emission in units of $\gamma=\gamma_a=\gamma_b$.
We consider two different parameter sets for resonator couplings: $g_a=g_b=1$ and $g_a=0.25$, $g_b=\sqrt{2}$. 
We find that when $g_a=g_b=1$ the probability for spurious emission
per bandwidth can be kept at $10^{-3}$ ($10^{-2}$) when $\gamma/2\pi =20$\,MHz ($100$\,MHz).
The noise in the case $g_a=0.25$, $g_b=\sqrt{2}$ is slightly higher.

The rate of these spontaneous emission events are also proportional to the real part of the impedance at $\delta\omega=\omega_{\rm J}-\omega_b$.
This rate can then be reduced by engineering an anti-resonance in the impedance at $\delta\omega$.
To numerically study the effect of an antiresonance, 
we modify the used impedance ${\rm Re}[Z_{\rm t}(\omega)]$ to
\begin{eqnarray}\label{eq:AntiResonance1}
{\rm Re}[Z_{\rm t}(\omega)]\rightarrow {\rm Re}[Z_{\rm t}(\omega)]\times\left[1- e^{-(\omega-\delta\omega)^2/2\Delta^2} \right] \, ,
\end{eqnarray}
with an antiresonance width $\Delta/2\pi=0.5$~GHz.
(The result depends only weakly on the chosen width, as long as $\Delta> \gamma_b$.)
In Fig.~\ref{fig:spontaneous}, we show the result when considering an antiresonance at $\delta \omega/2\pi=8-5=3$~GHz.
We get roughly an order of magnitude reduction in the rate for spontaneous emission.

Even when cascaded, such parasitic spontaneous emission is only photomultiplied
by the second stage or not at all. Parasitic emission, therefore, always produces lower photon numbers than a incoming photon. This means it is less likely to trigger a detection event. For example, in a cascaded setup with $n=3\times 3$, the most undesirable spontaneous emission
event is emission at the first amplification stage, which becomes multiplied by the second-state multiplier, and finally produces
three photons in the output. In this case, the probability
for triggering a click is $< 0.1$ if the vacuum dark count probability is set to $10^{-3}$ (see Fig.~\ref{fig:darkrate}). Here, by keeping the spontaneous emission
rate below $10^{-2}$, the spontaneous emission does not significantly increase the dark count rate.
In the discussed three-cavity setup,
the needed $E^{\rm in}_{\rm J}$ is also lower, for example, by a factor of one half for a rectangular bandwidth, see Fig.~\ref{fig:figureThreeCavity}, reducing the rate for spontaneous emission by a factor of four.
Note that the four times higher $E^{\rm out}_{\rm J}$ produces a higher parasitic spontaneous photon emission rate of the second stage multiplier. However, because these spontaneous photon emission events are not photomultiplied, they also do not significantly increase dark count rate.
We conclude that (in particular when using antiresonances) spontaneous emission can be reduced to a level where it does not dominate the
single-photon detection dark-count rate.

\subsection{Practical set of parameters and expected performance}
In order to summarize the results of this paper we give practical parameters for an experimental realization.
For photon tripling with resonator bandwidths $\gamma/2\pi=100$~MHz and identical couplings $g=1$ (meaning resonator characteristic impedances $Z=g^2R_{\rm Q}/\pi\approx 2.05$~k$\Omega$), we need $E_{\rm J}\approx 4.8$~$\mu$eV to have
the conversion probability as defined by filter function of Eq.~(\ref{eq:ResultBandwidth2}) with $n=3$.
This ideal conversion probability is reduced due to direct transmission (finite junction capacitance  $C_{\rm J}$) and thermal fluctuations of the bias voltage.
Keeping $C_{\rm J}< 1$~fF, the maximal conversion probability is reduced less than $3$~percent.
For $50$~$\Omega$ transmission line at $20$~mK thermal fluctuations reduce the conversion probability less than $10$~percent,
which can be reduced towards $1$~percent when decreasing the low-frequency impedance.
Spontaneous emission occurs with a rate $\sim 10^{-2}\times$~bandwidth, and can be reduced by engineering an anti-resonance.

To realize a single-photon detector through cascaded tripling and subsquent power detection,
we need three high-impedance resonators and two Josephson junctions (Sec.~\ref{sec:ThreeCavitySetup}).
The resonance frequencies have to be chosen carefully so that no unwanted resonances occur when voltage biasing.
The above analysis for spontaneous emission in the case of single-junction multiplier
is valid if we expand the used range of resonator frequencies, for example, from $5-7$~GHz to $5-9$~GHz.
When realizing the input and central cavity with couplings $g=1$ and the output cavity with $g=\sqrt{2}$, with bandwidths $\gamma_{\rm in/out}/2\pi=100$~MHz,
the first Josephson junction should have $E_{\rm J}\approx 2.4$~$\mu$eV and the second one $E_{\rm J}\approx 12$~$\mu$eV
to have conversion probability similar as in Fig.~\ref{fig:figureThreeCavity}(b) for $\epsilon^{\rm in}_3=1/2$ (rectangular shape).
Keeping $C_{\rm J}\sim 1$~fF the reduction in the conversion probability due to direct transmission is expected to stay within few percent also in this system and the effect of temperature is also similar. 
Finally, the power detection accuracy of created multi-photon Fock states can be made to be limited by vacuum fluctuations, depending on the chosen power threshold for a 'click', as described by Fig.~\ref{fig:darkrate}. One choice is the quantum efficiency $0.9$ which leads to dark-count rate $10^{-3}\times$~bandwidth.

\section{Conclusions and discussion}\label{sec:Conclusions}

In conclusion, we have shown that inelastic Cooper-pair tunneling can be used to deterministically convert propagating single microwave photons into multi-photon Fock states. Cascading two such multiplication stages, and reading them out using existing linear detection schemes, one can implement a microwave single photon detector with high detection efficiency, relatively low dark count rates and without dead time.
We also expect that the device is able to resolve photon numbers.
In comparison to photon-number doubling in parametric down conversion \cite{WallsMilburn},
the important difference is here that the energy absorbed from charge transport provides energy gain,
which allows for keeping the output photons in the same frequency range as the input photon.

There are also other intriguing physical properties of the created nonclassical microwave fields which could be exploited in other quantum applications.
The multi-photon Fock states are frequency entangled and can be highly bunched,
an outcome which could be interesting for quantum-information applications.
The creation of similar $N$-photon states (bundles) has been studied in cavity-QED systems~\cite{Bundles2014}.
Photon multiplication itself can be useful in nonlinear optical quantum computing~\cite{Langford2011}.
Moreover, similar multi-photon production between two superconducting resonators has also been studied recently as a versatile frequency converter~\cite{Anton2017}.
We also note that using this device backwards provides an engineered bath where multi-photon absorption is dominant.
This could be useful, for example, for \lq cat codes\rq~\cite{CatCodes} which encode an error-protected logical qubit in superpositions of coherent states.


\section*{Acknowledgments}
JL and MM acknowledge financial support from DFG Grant No.~MA 6334/3-1. 
DH, SJ, RA, FB, and MH acknowledge financial support from
Grenoble Nanosciences Foundation and from the European Research Council under the European
Union's Seventh Framework Programme (FP7/2007-2013) / ERC Grant agreement No 278203 -- WiQOJo.
GJ acknowledges financial support from the Swedish Research Council and the Knut and Alice Wallenberg Foundation.

\appendix*

\section{A: Heisenberg equations of motion}\label{sec:Heisenberg}
\begin{figure}
\includegraphics[width=\columnwidth]{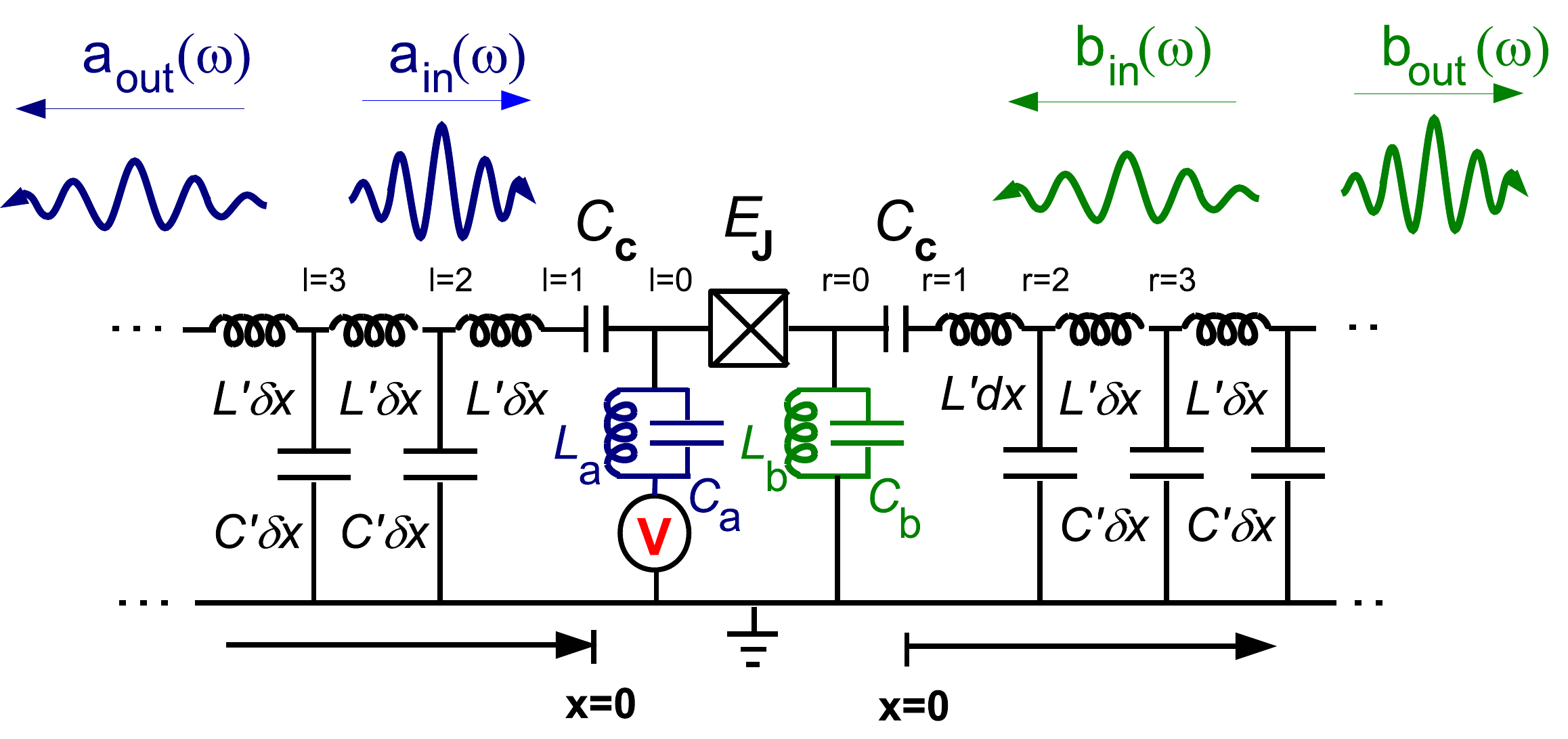}
\caption{Lumped-element model of the considered microwave circuit.}\label{fig:AppendixCircuit}
\end{figure}

In this Appendix, we derive the TL solution for propagating radiation, boundary conditions, and Heisenberg equations of motion used in the main part of the paper
starting from a continuous-mode treatment of the circuit shown in Fig.~\ref{fig:AppendixCircuit}.

\subsection{Lagrangian and Hamiltonian}
The total Lagrangian of the system shown in Fig.~\ref{fig:AppendixCircuit} can be decomposed as
\begin{equation}
{\cal L}={\cal L}_{\rm L}+{\cal L}_{\rm J} + {\cal L}_{\rm R}
\end{equation}
The left-hand side Lagrangian splits into ${\cal L}_{\rm L}={\cal L}_{\rm TL}+{\cal L}_{\rm a}$,
where the transmission-line part reads
\begin{eqnarray}
{\cal L}_{\rm TL}&=&\sum_{l \geq 2}^\infty\frac{\delta x C'\left(\dot\Phi_l+V\right)^2}{2}-\sum_{l\geq 2}^\infty\frac{(\Phi_l-\Phi_{l-1})^2}{2L'\delta x} \nonumber\\
&+&\frac{ C_{\rm c}\left( \dot\Phi_{1}-\dot\Phi_a \right)^2}{2}. 
\end{eqnarray}
Here $\Phi_l(t)$ is the magnetic flux of node $l$ and $\dot\Phi_l+V$ is the corresponding voltage.
This amounts to defining the magnetic flux (time integrated voltage) of the left-hand side transmission line with respect to $\Phi_V=Vt$.
The left-hand side oscillator part is
\begin{eqnarray}
{\cal L}_{a}&=&\frac{ C_a\dot\Phi_a^2}{2}-\frac{\Phi_a^2}{2L_a}.
\end{eqnarray}
Similarly for the oscillator $b$ and the right-hand side transmission line (with the value $V=0$).
The Josephson junction is described by the potential-energy term,
\begin{equation}
{\cal L_{\rm J}}=E_{\rm J}\cos\left( 2\pi \frac{\Phi_V+\Phi_{a}-\Phi_b}{\Phi_0}\right).
\end{equation}
Here $E_{\rm J}$ is the Josephson coupling energy and $\Phi_0=h/2e$ is the
flux quantum.

The above Lagrangian leads to the left-hand side Hamiltonian
\begin{eqnarray}
&&H_{\rm L}\equiv \sum_{i \in {\rm L}}\dot \Phi_iQ_i -{\cal L_{\rm L}} =\sum_{l=2}^{M} \frac{\left(Q_l-\delta x C'V\right)^2}{2\delta xC'} \\
&+&\sum_{l\geq 2}^N\frac{(\Phi_l-\Phi_{l-1})^2}{2L'\delta x}  + \frac{Q_a^2}{2C_{a}}+ \frac{ \Phi_a^2}{2L_a} +  \frac{Q_aQ_1}{C_a} +\frac{Q_1^2}{2C_{\rm s}},  \nonumber
\end{eqnarray}
where $Q_i=\partial{\cal L}/\partial \dot\Phi_i$ and $1/C_{\rm s}=1/C_{\rm c}+1/C_a$. 

For convenience, we can do a shift in the momentum variable and neglect the terms $\propto \delta x C'V$.
This does not change the Hamiltonian equations
\begin{eqnarray}
\frac{d Q_i}{dt}&=&-\frac{\partial H}{\partial \dot\Phi_i} \\
\frac{d \dot\Phi_i}{dt}&=&+\frac{\partial H}{\partial Q_i},
\end{eqnarray}
provided that $V$ is a constant.
Standard quantization means $\Phi_i\rightarrow \hat\Phi$ and $Q_i\rightarrow \hat Q_i$ with $[\hat\Phi_i,\hat Q_i]=i\hbar$.
Defining the normalized phase, $\hat \phi_i \equiv 2\pi \Phi_i/\Phi_0$, we have equivalently $[\hat\phi_i,\hat Q_i]=i2e$.

\subsection{Transmission line solution}
The Heisenberg equations of motion in the transmission line are ($l\geq 2$)
\begin{eqnarray}
\hat{\dot{ \Phi}}_i(t) &=& \frac{i}{\hbar}\left[\hat H,\hat\Phi_i \right]\\
\hat{\dot{ Q}}_i(t) &=& \frac{i}{\hbar}\left[\hat H,\hat Q_i \right].
\end{eqnarray}
These give us
\begin{eqnarray}
\hat{\dot{ \Phi}}_l(t) &=& \frac{\hat Q_l}{\delta x C'} \\
\hat{\dot{ Q}}_l(t) &=& \frac{ \hat\Phi_{l-1} + \hat\Phi_{l+1} -2\hat\Phi_l }{\delta x L'}.
\end{eqnarray}
In the continuum limit $\delta x\rightarrow 0$, the two equation lead to the Klein-Gordon equation,
\begin{equation}
\hat{\ddot{\Phi}}(x,t)=\frac{1}{L_i'C_i'}\frac{\partial^2\hat\Phi(x,t)}{\partial^2 x}.
\end{equation}
We can then establish a solution in the free space ($x<0$)
\begin{eqnarray}\label{AppendixEq:PropagatingField}
&&\hat\Phi(x,t)=\sqrt{\frac{\hbar Z_{0}}{4\pi}}\int_0^\infty\frac{d\omega}{\sqrt{\omega}}\times  \\
&&\left[ \hat a_{\rm  in}(\omega)e^{i(k_\omega x-\omega t)}+\hat a_{\rm out}(\omega)e^{i(-k_\omega x-\omega t)}+{\rm H.c.} \right]\, , \nonumber
\end{eqnarray}
where $Z_0=\sqrt{L'/C'}$ and $k_\omega=\omega\sqrt{L'C'}$.
Here, the operator $\hat a_{\rm in(out)}^{\dagger}(\omega)$ creates and the operator $\hat a_{\rm in(out)}(\omega)$ annihilates an
incoming (outgoing) propagating photon of frequency $\omega$.
We have the commutation relations
\begin{equation}\label{CommutationRelation}
\left[ \hat a_{\rm in}(\omega),\hat a_{\rm in}^{\dagger}(\omega') \right]=\delta(\omega-\omega')\, ,
\end{equation}
and similarly for the out-operators. The same derivation also applies for the propagating fields on
the right-hand side transmission line.

The last step is to take the narrow-bandwidth approximation, as described in Section~\ref{sec:TLOperators}.

\subsection{Resonator equations}
We introduce now the resonator creation and annihilation operators,
\begin{eqnarray}
 \hat \Phi_a &=& f\, \sqrt{\frac{\hbar}{2}}(\hat a+\hat a^\dagger)  \\
 \hat Q_a &=& \frac{i}{f}\, \sqrt{\frac{\hbar}{2}}(\hat a^\dagger - \hat a ).
\end{eqnarray}
Here for a free resonator the choice $f^2=Z_{LC}=\sqrt{L_a/C_a}$ diagonalizes the resonator Hamiltonian, and in this case
\begin{eqnarray}\label{eq:Phi}
\hat \phi=\frac{2\pi}{\Phi_0} \hat \Phi_a=\sqrt{\pi\frac{ Z_{LC}}{R_{\rm Q}}}\left( \hat a+ \hat a^\dagger\right).
\end{eqnarray}
$ \hat\phi=(2\pi/\Phi_0)\hat\Phi_a=\sqrt{\pi Z_{LC}/R_{\rm Q}}( \hat a+ \hat a^\dagger)$.
The resonance frequency has then the form $\omega_a=\sqrt{1/L_aC_a}$.
However, at this point we do not fix $Z_{LC}$ to this value, since the resonator capacitance will be normalized by the coupling capacitance $C_{\rm c}$,
as derived below. The form of Eq.~(\ref{eq:Phi}), however, stays the same, calculated with the renormalized capacitance.

At the resonator boundary ($l=1$) the Heisenberg equations of motion give
\begin{eqnarray}
\hat{\dot{ \Phi}}_1(t) &=&  \frac{\hat Q_a}{C_{a}}+ \frac{\hat Q_1}{C_{\rm s}} \label{eq:ResonatorBoundary1} \\
\hat{\dot{ Q}}_1(t) &=& \frac{ \hat \Phi_{2} -\hat \Phi_1 }{\delta x L'}\rightarrow  -\frac{1}{L'}\frac{\partial \hat \Phi(x=0,t)}{\partial x}. \label{eq:ResonatorBoundary2}
\end{eqnarray}
The derivative with respect to $x$ corresponds to the continuum limit $\delta x\rightarrow 0$.
A solution for the latter equation is
\begin{eqnarray}
\hat Q_1(t)&=&\sqrt{\frac{\hbar}{4\pi Z_0}}\int_0^\infty \frac{d\omega}{\sqrt{\omega}} \left[ \hat a_{\rm in}e^{-i\omega t} - \hat a_{\rm out}e^{-i\omega t} \right] + {\rm H.c.} \nonumber\\
&=&\sqrt{\frac{\hbar }{2\omega_{a}Z_{0}}}\left[\hat a_{\rm  in}(t)-\hat a_{\rm  out}(t)\right]+ {\rm H.c.} \, .
\end{eqnarray}
To proceed we now make an important observation.
In Eq.~(\ref{eq:ResonatorBoundary1}), the operator $\hat Q_1/C_{\rm s}$ is characterized by relative size
$\omega_c\equiv 1/C_{\rm s}Z_0$, whereas the time derivative of the phase $\hat{\dot{ \Phi}}_1(t)$ by size $\omega_a$.
Here, it is always the former term that will dominate (high cut-off frequency), and we can neglect the time derivative of the phase operator.
In this limit we get
\begin{eqnarray}
\hat a_{\rm out}(t)-\hat a_{\rm in}(t) &=& \alpha \hat a(t)  \\
\alpha &=& -i\frac{C_{\rm s}}{C_a}\sqrt{\frac{Z_0}{Z_{LC}}}\, \sqrt{\omega_a}.
\end{eqnarray}
To derive this we have used 
\begin{equation}
\hat Q_a=i\sqrt{\frac{\hbar}{2Z_{LC}}} \left[ \hat a^\dagger(t)-\hat a(t)\right].
\end{equation}

At the junction, the effective Hamiltonian to be used in the Heisenberg equations of motion has the form
\begin{widetext}
\begin{eqnarray}
&&\hbar\left[  \frac{Z_{LC}}{4L_a}\left(\hat a+\hat a^\dagger\right)^2 - \frac{1}{4Z_{LC}C_a}\left(\hat a-\hat a^\dagger\right)^2  \right]+\hat H_{\rm J}+i\frac{1}{C_a}\hat Q_1\sqrt{\frac{\hbar}{2Z_{LC}}}\left[ \hat a^\dagger -\hat a  \right]=  \\
=&&\hbar\left[  \frac{Z_{LC}}{4L_a}\left(\hat a+\hat a^\dagger\right)^2 - \frac{1}{4Z_{LC}C_a}\left(\hat a-\hat a^\dagger\right)^2  \right]+\hat H_{\rm J}+i\frac{1}{C_a}\left[ C_{\rm s}\hat{\dot{\Phi}}(t,0) -i\frac{C_{\rm s}}{C_a}\sqrt{\frac{\hbar}{2Z_{LC}}}(\hat a^\dagger-\hat a) \right]\sqrt{\frac{\hbar}{2Z_{LC}}}\left[ \hat a^\dagger -\hat a  \right], \nonumber
\end{eqnarray}
\end{widetext}
where we used Heisenberg Eq.~(\ref{eq:ResonatorBoundary1}) to eliminate $\hat Q_1$.
The last term inside the second parentheses contributes to the effective capacitance of the resonator, changing it to $C_{\rm p}=C_a+C_{\rm s}$.
More rigorously: the choice $Z_{LC}=\sqrt{L_a/C_{\rm p}}$ leads to the quadratic resonator part $(\hbar/\sqrt{L_aC_{\rm p}})\hat a^\dagger \hat a=\hbar\bar\omega_a\hat a^\dagger \hat a$.
Using the relation
\begin{equation}
\hat{\dot{\Phi}}(0,t)=-i\sqrt{\frac{\hbar Z_0\omega_a}{2}} \left[ \hat a_{\rm  in}(t)+\hat a_{\rm out}(t) \right],
\end{equation}
the Heisenberg equations take the form
\begin{eqnarray}
&&\hat{\dot {a}}(t) = \\
&&-i\bar\omega_a \hat a(t) +\frac{\alpha}{2}\left[ \hat a_{\rm in}+\hat a_{\rm out} \right] +\frac{i}{\hbar}\left[ \hat H_{\rm J}, \hat a \right] \,. \nonumber
\end{eqnarray}
Using $\hat a_{\rm out}(t)-\hat a_{\rm in}(t)=\alpha \hat a(t)$ and defining $\gamma_a=\vert \alpha\vert^2$ one arrives in the equation of motion
\begin{eqnarray}
&&\hat{\dot {a}}(t) = \\
&&-i\bar\omega_a \hat a(t) -\frac{\gamma_a}{2}\hat a(t) -i\sqrt{\gamma_a} \hat a_{\rm in} +\frac{i}{\hbar}\left[ \hat H_{\rm J}, \hat a \right] \,. \nonumber
\end{eqnarray}
We would like to express the boundary condition and the equation of motion in a form used often in the literature.
We do this by
redefining the phase of the operators $\hat a_{\rm in}\rightarrow -i \hat a_{\rm in}$ and $\hat a_{\rm out}\rightarrow i \hat a_{\rm out}$,
which leads to
\begin{eqnarray}
\hat{\dot{ a}}(t) &=& \\
&-&i\bar\omega_a \hat a(t) -\frac{\gamma_a}{2}\hat a(t) +\sqrt{\gamma_a} \hat a_{\rm in} +\frac{i}{\hbar}\left[ \hat H_{\rm J}, \hat a \right]\nonumber \\
\sqrt{\gamma_a}\hat a(t)&=&\hat a_{\rm in}(t)+\hat a_{\rm out}(t)\, .
\end{eqnarray}
Similar Heisenberg equations can also be derived  for the right-hand side transmission-line operators.



\section{B: Single-to-multi-photon scattering matrix}
In this Appendix, we derive the single-to-multi-photon scattering matrix given in the main part of the article.
Our goal is to determine the amplitude (scattering matrix)
\begin{eqnarray}
A&=&\left\langle 0 \left\vert \hat  b_{\rm out}(p_1) \hat b_{\rm out}(p_2)\ldots \hat b_{\rm out}(p_n)  \hat a^\dagger_{\rm in}(k) \right\vert 0 \right\rangle ,
\end{eqnarray}
with the help of resonator boundary conditions and Heisenberg equations of motion.

\subsection{Decoupled resonators ($E_{\rm J}=0$)}\label{sec:UncoupledModes}
In the following calculation, we will need to know the solution for the problem when $\epsilon_{\rm I}=0$ ($E_{\rm J}=0$), i.e.,
when resonator $a$ is decoupled from resonator $b$. Considering incoming radiation from the transmission line $a$,
we only need to solve the equation
\begin{equation}
\dot {\hat{a}}(t)= -i\omega_a\hat a(t)  - \frac{\gamma_a}{2}\hat a(t)+\sqrt{\gamma_a}\hat a_{\rm in}(t).
\end{equation}
A Fourier transformation gives
\begin{equation}
-i\omega \hat a(\omega)=-i\omega_a\hat a(\omega) - \frac{\gamma_a}{2}\hat a(\omega)+\sqrt{\gamma_a}\hat a_{\rm in}(\omega).
\end{equation}
The solution for the resonator field is then
\begin{equation}
\hat a(\omega)=\frac{\sqrt{\gamma_a}}{i(\omega_a-\omega)+\frac{\gamma_a}{2}}\hat a_{\rm in}(\omega),
\end{equation}
whereas the out-field has the form
\begin{equation}\label{eq:InOut0}
\hat a_{\rm out}(\omega)= \frac{\frac{\gamma_a}{2}-i(\omega_a-\omega)}{\frac{\gamma_a}{2}+i(\omega_a-\omega)}\hat a_{\rm in}(\omega).
\end{equation}
Similar relations are also valid for the propagating fields in the transmission line $b$.
This result states that all incoming radiation will be reflected with a specific phase shift.
At resonance ($\omega=\omega_a$) we have $\hat a(\omega)=(2/\sqrt{\gamma_a})\hat a_{\rm in}(\omega)$,
which means that in the relation $\hat a_{\rm out}(\omega)=\sqrt{\gamma_a}\hat a(\omega)-\hat a_{\rm in}(\omega)$
the contribution from the cavity is exactly twice the incoming field. 
On the other hand, in the case $\epsilon_{\rm I}\neq 0$,
we aim for the opposite situation, where these two contributions cancel each other and there will be no reflection.

\subsection{Scattering matrix in the case $n=2$}
We continue by considering in detail the case $n=2$ and then describe the generalization to arbitrary $n$.
Here, we evaluate the scattering element
\begin{eqnarray}
A&=&\left\langle 0 \left\vert  \hat b_{\rm out}(p_1) \hat b_{\rm out}(p_2)  \hat a^\dagger_{\rm in}(k) \right\vert 0 \right\rangle \\
&=&\left\langle 0 \left\vert  \hat b_{\rm out}(p_1) \hat b_{\rm out}(p_2) \left[ \sqrt{\gamma_a}\hat a^\dagger(k)-\hat a^\dagger_{\rm out}(k)  \right] \right\vert 0 \right\rangle\, .  \nonumber
\end{eqnarray}
Since the out fields of different modes (by definition) need to commute, we must have
\begin{eqnarray}
&&\left\langle 0 \left\vert  \hat b_{\rm out}(p_1) \hat b_{\rm out}(p_2) \hat a^\dagger_{\rm out}(k)  \right\vert 0 \right\rangle\\
&&=\left\langle 0 \left\vert \hat b_{\rm out}(p_1) \hat a^\dagger_{\rm out}(k)  \hat b_{\rm out}(p_2)   \right\vert 0 \right\rangle=0\, . \nonumber
\end{eqnarray}
Therefore,
\begin{eqnarray}
&&\left\langle 0 \left\vert  \hat b_{\rm out}(p_1) \hat b_{\rm out}(p_2)  \hat a^\dagger_{\rm in}(k) \right\vert 0 \right\rangle \\
&&= \sqrt{\gamma_a} \left\langle 0 \left\vert  \hat b_{\rm out}(p_1) \hat b_{\rm out}(p_2)  \hat a^\dagger(k) \right\vert 0 \right\rangle \,. \nonumber
\end{eqnarray}
For simplicity of Heisenberg equations of motion, we define now
\begin{eqnarray}
\bar\epsilon &=&-i\frac{\epsilon_{\rm I}}{(2\pi)^{(n-1)/2}}\, .
\end{eqnarray}
We then get
\begin{widetext}
\begin{eqnarray}
A=\sqrt{\gamma_a} \left\langle 0 \left\vert  \hat b_{\rm out}(p_1) \hat b_{\rm out}(p_2)  \hat a^\dagger(k) \right\vert 0 \right\rangle =\frac{\gamma_a}{F_a(\omega_k)}\left\langle 0 \left\vert \hat b_{\rm out}(p_1) \hat b_{\rm out}(p_2)  \left[ \hat a^\dagger_{\rm out}(k) +\frac{\bar\epsilon}{\sqrt{\gamma_a}}\int d\omega'\hat b^\dagger(\omega')\hat b^\dagger(\omega_k+\omega_{\rm J}-\omega')\right] \right\vert 0 \right\rangle \, ,\nonumber
\end{eqnarray}
\end{widetext}
where in the second form we used Eq.~(\ref{eq:Heienberg1}).
As the first term (inside the square brackets) again gives no contribution, we must have
\begin{eqnarray} \label{eq:general1}
A&=&\frac{\gamma_a}{F_a(\omega_k)} \frac{\bar\epsilon}{\sqrt{\gamma_a}} \int d\omega'  \\
&\times&\left\langle 0 \left\vert  \hat b_{\rm out}(p_1) \hat b_{\rm out}(p_2) \hat b^\dagger(\omega')\hat b^\dagger(\omega_k+\omega_{\rm J}-\omega') \right\vert 0 \right\rangle\, . \nonumber
\end{eqnarray}
We continue by exploiting the Heisenberg equation
\begin{eqnarray}
&&-F_b(\omega)\hat b^\dagger(\omega)= -\sqrt{\gamma_b}\hat b^\dagger_{\rm out}(\omega)\label{eq:general1}\\
&& + 2\times\bar\epsilon^* \int d\omega'\hat a^\dagger(\omega')\hat b(\omega'+\omega_{\rm J}-\omega)\,.  \nonumber
\end{eqnarray}
The factor 2 comes from the factor $n$ in the boundary condition. When applying $b$ to the ground state we obtain zero
[since $b=(b_{\rm in}+ b_{\rm out})/\sqrt{\gamma_b}$ and the ground state has no incoming or outgoing photons].
Therefore,
\begin{widetext}
\begin{eqnarray}
&&A=\frac{\gamma_a}{F_a(\omega_k)}\frac{\bar\epsilon}{\sqrt{\gamma_a}} \int d\omega' \left\langle 0 \left\vert \hat b_{\rm out}(p_1) \hat b_{\rm out}(p_2)  \hat b^\dagger(\omega')\frac{\sqrt{\gamma_b}}{F_b(\omega_k+\omega_{\rm J}-\omega')}\hat b^\dagger_{\rm out} (\omega_k+\omega_{\rm J}-\omega') \right\vert 0 \right\rangle.  
\end{eqnarray}
\end{widetext}
In order to evaluate application by $b^\dagger(\omega')$, we again make use of the Heisenberg equation~(\ref{eq:general1}).
Consider first the term not proportional to $\bar\epsilon^*$, i.e., another multiplication by $b^\dagger_{\rm out}$.
Using
\begin{eqnarray}
&&\left\langle 0\vert b(f_1)b(f_2) b^\dagger(f_3)b^\dagger(f_4)\vert 0 \right\rangle \label{eq:general2} \\
&&=\delta(f_1-f_3)\delta(f_2-k_4)+\delta(f_1-f_4)\delta(f_2-f_3)\,, \nonumber
\end{eqnarray}
we get for this term (we name it $A_0$),
\begin{eqnarray}\label{eq:Generalization1}
&&A_0=\bar\epsilon\frac{\sqrt{\gamma_a}}{F_a(\omega_k)}\times     \label{eq:BarA}    \\
&&\left[ \frac{\sqrt{\gamma_b}}{F_b({p_1})}\frac{\sqrt{\gamma_b}}{F_b(\omega_k+\omega_{\rm J}-{p_1})} + \frac{\sqrt{\gamma_b}}{F_b({p_2})}\frac{\sqrt{\gamma_b}}{F_b(\omega_k+\omega_{\rm J}-{p_2})} \right]\nonumber \\
&& \times\delta(\omega_k+\omega_{\rm J}-{p_1}-{p_2}) = \bar A_0\delta(\omega_k+\omega_{\rm J}-{p_1}-{p_2})\,.  \nonumber
\end{eqnarray}
The two terms inside the square brackets of Eq.~(\ref{eq:Generalization1}) are equal. This is the leading-order solution in $\bar\epsilon$.

The second contribution to $A$ accounts for the non-perturbative limit. We need to evaluate
\begin{widetext}
\begin{eqnarray}
&& -\int d\omega'  f(\omega')\int d\omega''\left\langle 0 \left\vert  \hat b_{\rm out}(p_1) \hat b_{\rm out}(p_2)   \hat a^\dagger(\omega'')\hat b(\omega''+\omega_{\rm J}-\omega')   \hat b^\dagger_{\rm out} (\omega_k+\omega_{\rm J}-\omega') \right\vert 0 \right\rangle=\nonumber \\
&&-\int d\omega' f(\omega')\int d\omega''\left\langle 0 \left\vert  \hat b_{\rm out}(p_1) \hat b_{\rm out}(p_2)  \hat a^\dagger(\omega''){\cal I}  \hat b(\omega''+\omega_{\rm J}-\omega')  \hat b^\dagger_{\rm out} (\omega_k+\omega_{\rm J}-\omega') \right\vert 0 \right\rangle \label{eq:OtherContribution}. \label{eq:Generalization2}
\end{eqnarray}
\end{widetext}
Here we have defined
\begin{eqnarray}
f(\omega')=2\vert\bar\epsilon\vert^2\frac{\sqrt{\gamma_a}}{F_a(\omega_k)}  \frac{\sqrt{\gamma_b}}{F_b(\omega_k+\omega_{\rm J}-\omega')}\frac{1}{F_b(\omega')}\, .
\end{eqnarray}
In the second form we have also inserted identity operator $\cal I$
between $a^\dagger$ and $b$.

A crucial step here is based on the observation:
only insertion ${\cal I}\rightarrow \vert 0 \rangle\langle 0 \vert $ gives nonzero contribution.
Similar property has also been used to evaluate scattering properties on a two-level system~\cite{Fan2010} and is possible due to
photon-number conservation (here of the specific form $n_a+n_b/n={\rm constant}$).
Eq.~(\ref{eq:OtherContribution}) becomes then a product of two amplitudes.
The left-hand side amplitude is proportional to $A$ and the right-hand side amplitude
measures scattering of single incoming photon from side $b$. Single incoming photon from side $b$
has no change but to reflect at the junction in a way described by the solution for $\epsilon=0$.  This solution is
derived in Sec.~\ref{sec:UncoupledModes}, where we get
\begin{eqnarray}
\left\langle 0 \left\vert  \hat b(p)   \hat b^\dagger_{\rm out} (k) \right\vert 0 \right\rangle =\frac{\sqrt{\gamma_b}}{F_b^*(\omega_k)}\delta(\omega_p-\omega_k).
\end{eqnarray}
Therefore, the term in Eq.~(\ref{eq:OtherContribution}) can be rewritten in the form
\begin{eqnarray}
&& -\frac{A}{\sqrt{\gamma_a}}\int d\omega'  f(\omega')  \frac{\sqrt{\gamma_b}}{F^*_b(\omega_k+\omega_{\rm J}-\omega')} .
\end{eqnarray}
We then continue by evaluating the integration $\int d\omega'$ explicitly, which is done over the function
\begin{eqnarray}
&&  A\frac{-2\vert\epsilon\vert^2}{F_a(\omega_k)} \frac{\sqrt{\gamma_b}}{F_b(\omega_k+\omega_{\rm J}-\omega')}  \frac{1}{F_b(\omega')}  \frac{\sqrt{\gamma_b}}{F^*_b(\omega_k+\omega_{\rm J}-\omega')}  \nonumber \\
&& =    -\frac{2\vert\epsilon\vert^2 A}{F_a(\omega_k)}\ \ \left \vert \frac{\sqrt{\gamma_b}}{\gamma_b/2+i(\omega_k+\omega_{\rm J}-\omega'-\omega_b)} \right\vert^2 \nonumber\\
&&\times \frac{1}{\gamma_b/2+i(\omega_b-\omega')}\, .
\end{eqnarray}
Let us mark $\delta\omega=\omega_k+\omega_{\rm J}-2\omega_b$ (which is ideally zero).
An analytical integration is possible and leads to the relation
\begin{eqnarray}
A= A_0 -  A\frac{1}{F_a(\omega_k)} \frac{4\pi\vert\bar\epsilon\vert^2 }{\gamma_b+i\delta\omega}.
\end{eqnarray}
This means
\begin{eqnarray}
A&=& \frac{A_0}{1+a}  \\
 a&=&4\pi\frac{\vert\bar\epsilon\vert^2 }{F_a(\omega_k)(\gamma_b+i\delta\omega)}\nonumber\\
 &=& 2\frac{\vert\epsilon_{\rm I}\vert^2 }{F_a(\omega_k)(\gamma_b+i\delta\omega)}.
\end{eqnarray}
where in the last form we went back to the original definition of $\vert\epsilon_{\rm I}\vert=\sqrt{2\pi}\vert\bar\epsilon\vert$.
The amplitude $A_0$ was was derived above,
\begin{widetext}
\begin{eqnarray}
&&A_0=2\times \bar\epsilon\frac{\sqrt{\gamma_a}}{F_a(\omega_k)} \frac{\sqrt{\gamma_b}}{F_b(\omega_{p_1})}\frac{\sqrt{\gamma_b}}{F_b(\omega_k+\omega_{\rm J}-\omega_{p_1})}  \delta(\omega_k+\omega_{\rm J}-\omega_{p_1}-\omega_{p_2}). 
\end{eqnarray}
\end{widetext}
For the ideal case $\omega_k=\omega_a$ and $\delta\omega=0$ we get
\begin{eqnarray}
A&=& \frac{1}{1+a}\times 4\frac{\bar\epsilon}{\sqrt{\gamma_a}}\frac{\gamma_b}{(\omega_b-\omega_{p1})^2+\gamma_b^2/4}\\
&\times &\delta(2\omega_b-\omega_{p_1}-\omega_{p_2}) \nonumber  \\
 a&=&\frac{4\vert\epsilon_{\rm I}\vert^2 }{\gamma_a\gamma_b}\, .
\end{eqnarray}

To find the multiplication probability, we evaluate the photon number on side $b$.
This means evaluating
\begin{eqnarray}\label{eq:generalisationProbability0}
P=\int d\omega\int d\omega'\langle 1_a \vert \hat b^\dagger_{\rm out}(\omega) \hat b_{\rm out}(\omega') \vert 1_a\rangle.
\end{eqnarray}
The trick here is to insert a single-$b$-side-photon state in between the two operators (based on the same observation as
made when calculating Eq.~[\ref{eq:OtherContribution})],
\begin{eqnarray}
P&=&\int d\omega\int d\omega' \int d\omega'' \label{eq:generalisationProbability} \\
&\times& \langle 1_a \vert \hat b^\dagger_{\rm out}(\omega)\vert 1_{b\,\omega''} \rangle\langle 1_{b\,\omega''} \vert \hat b_{\rm out}(\omega') \vert 1_a\rangle\, , \nonumber
\end{eqnarray}
which means that the photon number has the form
\begin{eqnarray}
P=\frac{1}{(1+a)^2}\int d\omega_{p_1}\vert \bar A_0 (\omega_{p_1},\omega_k-\omega_{p_1}) \vert^2.
\end{eqnarray}
Here $\bar A_0(\omega_{p_1},\omega_{p_2})$ was defined to be the same as $A_0$ but without the delta-function, Eq.~(\ref{eq:BarA}).
The integration is over a product of two Lorenzian functions and can again be performed analytically.
For the ideal case $\omega_k=\omega_a$ and $\delta\omega=0$ we get
\begin{eqnarray}
P=2\times \frac{4a}{(1+a)^2},
\end{eqnarray}
which is the final result.

\subsection{Scattering matrix for general $n$}
Let us discuss now how the previous derivation is modified in the case of general $n$. In this situation,
Eq.~(\ref{eq:general2}) gets generalized to $n!$ identical contributions,
leading to the leading-order amplitude
\begin{eqnarray}
A_0&=& n!\times \bar\epsilon\frac{\sqrt{\gamma_a}}{F_a(\omega_k)} \frac{\sqrt{\gamma_b}}{F_b({p_1})}\ldots \frac{\sqrt{\gamma_b}}{F_b({p_n})}\\
&\times&\delta(\omega_k+\omega_{\rm J}-{p_1}-\ldots-{p_n})\,. \nonumber
\end{eqnarray}

Eq.~(\ref{eq:general1}) includes a factor $n$ instead of the factor 2.
Applying this equation to solve Eq.~(\ref{eq:Generalization2}), we  get a higher-order contribution in $\bar \epsilon$ only 
when the last of the creations by $\hat b^\dagger$ is replaced by $n\times \hat a^\dagger (\tilde \omega_1) \hat b(\tilde{\omega_2}) \ldots \hat b(\tilde\omega_{n-1})\hat b(\tilde\omega_1 +\omega_{\rm J}-\omega_1-\tilde\omega_2-\ldots-\tilde\omega_{n-1})$.
The other terms (on the right-hand side of this) contribute with the zeroth-order term,
 $\hat b^\dagger_{\rm out}(\omega_2)\ldots \hat b^\dagger_{\rm out}(\omega_{n-1})\hat b^\dagger_{\rm out}(\omega_k+\omega_{\rm J}-\omega_1-\ldots-\omega_{n-1})$.
This leads to the general form of the function
\begin{eqnarray}
&&f(\omega')\rightarrow f(\omega_1,\ldots,\omega_{n-1})= n\times\vert\bar\epsilon\vert^2\times \\
&&\frac{\sqrt{\gamma_a}}{F_a(\omega_k)} \frac{1}{F_b(\omega_1)}\frac{\sqrt{\gamma_b}}{F_b(\omega_2)}\ldots \frac{\sqrt{\gamma_b}}{F_b(\omega_k+\omega_{\rm J}-\omega_1-\ldots\omega_{n-1})}\,. \nonumber
\end{eqnarray}

Eq.~(\ref{eq:OtherContribution}) is then again a product of two amplitudes.
The left-hand side amplitude is proportional to $A$ and the right-hand side amplitude
measures scattering of $n-1$ incoming photons from side $b$.
The (left-hand side) factor $A$ has now been evaluated with respect to final state $\tilde \omega_1$.
Again, $n-1$ incoming photon from side $b$
has no change but to reflect at the junction in a way described by the solution for $\epsilon=0$.
We can then evaluate the (right-hand side) expectation value
\begin{widetext}
\begin{eqnarray}
&&E=\left\langle  \hat b(\tilde{\omega_2}) \ldots \hat b(\tilde\omega_{n-1})\hat b(\tilde\omega_1 +\omega_{\rm J}-\omega_1-\tilde\omega_2-\ldots-\tilde\omega_{n-1})  \times \hat b^\dagger_{\rm out}(\omega_2)\ldots \hat b^\dagger_{\rm out}(\omega_{n-1})\hat b^\dagger_{\rm out}(\omega_k+\omega_{\rm J}-\omega_1-\ldots-\omega_{n-1})  \right\rangle, \nonumber
\end{eqnarray}
\end{widetext}
by using the relation between $\hat b$ and $\hat b_{\rm out}$ obtained for $\bar \epsilon=0$
(Section~\ref{sec:UncoupledModes}), which gives
\begin{eqnarray}
&&\int d\tilde\omega_1\ldots d\tilde\omega_{n-1} E=  \\
&&(n-1)!\frac{\sqrt{\gamma_b}}{F_b^*(\omega_2)}\ldots \frac{\sqrt{\gamma_b}}{F_b^*(\omega_k+\omega_{\rm J}-\omega_1-\ldots-\omega_{n-1})}\,, \nonumber
\end{eqnarray}
and $\tilde\omega_1=\omega_k$ in the (left-hand side) matrix element corresponding to amplitude $A$.
The last step is then to evaluate the integral
\begin{widetext}
\begin{eqnarray}
\frac{A}{F_a(\omega_k)}\vert \bar\epsilon\vert^2 n\times(n-1)!\int d\omega_1\ldots d\omega_{n-1} \left\vert\frac{\sqrt{\gamma_b}}{F_b(\omega_2)}\right\vert^2\ldots \left\vert \frac{\sqrt{\gamma_b}}{F_b(\omega_k+\omega_{\rm J}-\omega_1-\ldots-\omega_{n-1})}\right\vert^2\times \frac{1}{F_b(\omega_1)} ,
\end{eqnarray}
\end{widetext}
The evaluation can again be done analytically and  gives the result
\begin{eqnarray}
\frac{A\vert \bar\epsilon\vert^2}{F_a(\omega_k)}\frac{2n!(2\pi)^{n-1}}{n\gamma_b-2i\delta\omega}.
\end{eqnarray}
This leads us to the result in the ideal case $\delta\omega=0$, $\omega_k=0$,
\begin{eqnarray}
A=\frac{A_0}{1+a}, \,\,\,\,\, a=4\frac{(n-1)!\vert \epsilon_{\rm I}\vert^2}{\gamma_a\gamma_b}.
\end{eqnarray}
In the more general form we have
\begin{eqnarray}
a=\frac{1}{F_a(\omega_k)}\frac{2n!\vert \epsilon_{\rm I}\vert^2}{n\gamma_b-2i\delta\omega}.
\end{eqnarray}

The photon number on side $b$ is evaluated similarly as in the case $n=2$, by inserting a state
$(1/\sqrt{(n-1)}!)\int d\omega_1 \ldots d\omega_{n-1} \hat b^{\dagger}_{\rm out}(\omega_1)\ldots \hat b^{\dagger}_{\rm out}(\omega_{n-1}) \vert 0\rangle$
between the operators in the expectation value of Eq.~(\ref{eq:generalisationProbability0}).
The result for the photon number on side $b$ for general $n$ agrees with the result from the linearization method described below.

\section{C: Evaluation of the second-order coherence}
\subsection{Definition of wavepackets}
The second-order coherence $g^{2}(\tau)$ compares the probability of measuring one photon to measuring two photons
within certain time difference $\tau$. The result tells how photons appear in a detector: randomly $g^{2}(\tau)=1$, bunched $g^{2}(\tau)>1$, or antibunched $g^{2}(\tau)<1$~\cite{Loudon,WallsMilburn}.
To evaluate this for propagating multi-photon Fock states, we need to introduce a finite-width wavepacket.
This is since (i) single photons states have in reality finite widths and (ii) the result for
infinitely long wavepackets (in time and space) is infinity, as we show below.

Consider first a Gaussian waveform,
\begin{eqnarray}
&&\xi(\omega)=\\
&&\left(\frac{1}{2\Delta^2}\right)^{1/4}\exp\left[ -i(\omega_a-\omega)t_0-\frac{\left(\omega_a-\omega\right)^2}{4\Delta^2}  \right] \, . \nonumber
\end{eqnarray}
Here $t_0$ is the time at which the peak of pulse passes the detection point.
In the following, we assume $\Delta \ll \gamma_a$, i.e. the scattering matrix $A$ (and factors $\alpha$) can be treated as a constant when integrating over
the input frequency $\omega_{\rm in}$. 
This means this degree of freedom can be integrated out from the expressions, leading to the contribution
\begin{eqnarray}
\xi(t)&=&\int d\omega e^{-i\omega t}\xi(\omega)=\sqrt{2\pi}\left(\frac{2\Delta^2}{\pi}\right)^{1/4}\\
&\times&\exp\left[ -i\omega_at-\Delta^2\left(t_0-t\right)^2  \right] \, . \nonumber
\end{eqnarray}
We have the normalization $\int d\omega\vert\xi(\omega)\vert^2=\int d t\vert\xi(t)\vert^2/2\pi=1$.

We can then define a propagating single-photon state~\cite{Loudon}
\begin{eqnarray}
\vert 1_\xi\rangle=\int d\omega\xi(\omega)\hat b^\dagger\vert 0\rangle \, .
\end{eqnarray}
We have $\langle 1_\xi \vert 1_\xi\rangle=1$ and the photon number
\begin{eqnarray}
\langle 1_\xi\vert\hat n \vert 1_\xi\rangle&=&\langle 1_\xi\vert \int d\omega\hat b^\dagger(\omega)\hat b(\omega) \vert 1_\xi\rangle\\
&=&\int d\omega\vert\xi(\omega)\vert^2=1 \, . \nonumber
\end{eqnarray}
Note that in this case the photon number operator $\hat n$ is defined as diagonal in frequencies, whereas earlier we used a non-diagonal form, in
Eq.~(\ref{eq:Probability}). The difference originates in that earlier we worked with single-photon creation operators, rather than single-photon wavepackets, and the two treatments can be shown to be equivalent.

Consider now a narrow-bandwidth wavepacket, $\Delta\ll \gamma$, so that practically that all frequency components of an incoming single-photon state
with frequency $\omega_a$ are converted with probability $1$. The out state that is consistent with the scattering matrix, Eq.~(\ref{eq:ResultEntanglement}), has the form
\begin{widetext}
\begin{eqnarray}\label{eq:SolutionApp}
\vert \rm out\rangle&=&\int d\omega_{\rm in}\xi(\omega_{\rm in})\frac{1}{n!}\int d\omega_1\ldots d\omega_{n}   \hat b^\dagger({\omega_1}) \ldots \hat b^\dagger(\omega_{n}) B(\omega_1,\ldots\omega_{n-1})\delta(\omega_{\rm in}+\omega_{\rm J}-\omega_1-\ldots -\omega_n)\vert 0 \rangle  \\
B&=&n!\frac{\sqrt{\gamma_b}}{\sqrt{(n-1)!}}\frac{1}{(2\pi)^{(n-1)/2}} \frac{\epsilon_n}{1+\vert\epsilon_n\vert^2}\beta(\omega_1)\ldots\beta(\omega_n)   \, \, .
\end{eqnarray}
\end{widetext}
The function $\beta(\omega)=\sqrt{\gamma_b}/[i(\omega_b-\omega)+\gamma_b/2]$. We put here $\epsilon_n=1$ (perfect transmission).
It can be shown that the normalization condition $ \langle \rm out\vert \rm out\rangle=1$ is here equivalent with the condition
\begin{eqnarray}
&&\int d\omega_1\ldots \int d\omega_{n-1} \\
&&\vert B(\omega_1,\omega_2,\ldots,\omega_{n-1},\omega_{\rm in}+\omega_{\rm J}-\omega_1-\ldots-\omega_{n-1})\vert^2 \, . \nonumber\\
&&=n!  \, . \nonumber 
\end{eqnarray}
Applying this for the obtained amplitude (for the case $\epsilon_n=1$) we confirm that this is indeed the case for the presented solution.
We are then ready to evaluate the first and second-order coherences of such pulse fields.

\subsection{Second-order coherence}
The (unnormalized)  first-order coherence for propagating fields can be defined as
\begin{eqnarray}\label{eq:AppCEq1}
G^{(1)}(\tau,t)&\equiv& \frac{\hbar Z_0}{4\pi}\int d\omega \int d\omega'  \sqrt{\omega\omega'}e^{i\omega(t+ \tau)}e^{-i\omega' t}\nonumber \\
&\times&\left\langle \hat b_{\rm out}^\dagger(\omega) \hat b_{\rm out}(\omega') \right\rangle   \, . 
\end{eqnarray}
The (unnormalized) second-order coherence (at the photomultiplier $x=0$) is defined similarly,
\begin{widetext}
\begin{eqnarray}\label{eq:AppCEq1}
G^{(2)}(\tau,t)&\equiv& \left( \frac{\hbar Z_0}{4\pi} \right)^2 \int d\omega \int d\omega'\int d\omega'' \int d\omega''' \nonumber\\
&\times& \sqrt{\omega\omega'\omega''\omega'''}  e^{i\omega t}  e^{i\omega'(t+ \tau)} e^{-i\omega''(t+ \tau)}e^{-i\omega''' t}   \left\langle \hat b_{\rm out}^{\dagger}(\omega)\hat  b_{\rm out}^{\dagger}(\omega')\hat  b_{\rm out}(\omega'') \hat b_{\rm out}(\omega''') \right\rangle \, .
\end{eqnarray}
\end{widetext}
In the following, we use the solution of Eq.~(\ref{eq:SolutionApp}) for the out field.
For general $n$, the (equal-time) first-order coherence gets a simple form
\begin{eqnarray}\label{eq:AppCEq3}
G^{(1)}(0,t)&=& n\frac{\hbar Z_0}{4\pi}\omega_b\vert\xi(t)\vert^2\\
&=&n \frac{\hbar Z_0}{4\pi}\omega_b 2\pi\sqrt{\frac{2\Delta^2}{\pi}}\exp\left[ -2\Delta^2(t-t_0)^2  \right]\, , \nonumber
\end{eqnarray}
where we have used the narrow-bandwidth approximation. Similarly, we evaluate the photon flux 
\begin{eqnarray}\label{eq:AppCEq4}
F(t)&=& \frac{1}{2\pi}\int d\omega \int d\omega' \left\langle \hat b_{\rm out}^\dagger(\omega) \hat b_{\rm out}(\omega') \right\rangle \nonumber \\
&=&n \sqrt{\frac{2\Delta^2}{\pi}} \exp\left[ -2\Delta^2(t-t_0)^2  \right]  \, .
\end{eqnarray}
The total amount of photons in the transmission line is then consistently (integration over $t$) $n$.

Inserting the solution of Eq.~(\ref{eq:SolutionApp}) to the second-order coherence we find ($\tau>0$)
\begin{eqnarray}\label{eq:AppCEq1}
G^{(2)}(\tau,t)&=& \left( \frac{\hbar Z_0}{4\pi} \right)^2\omega_b^2 \pi\gamma_b n(n-1)e^{-\gamma\tau} \\
&\times& 2\pi\sqrt{\frac{2\Delta^2}{\pi}} \exp\left[ -2\Delta^2(t-t_0)^2  \right] \, .  \nonumber
\end{eqnarray}
This gives for the normalized second-order coherence
\begin{eqnarray}
g^{(2)}(\tau,t)&=&\frac{G^{(2)}(\tau,t)}{\vert G^{(1) }(0,t)\vert^2}= \left( 1- \frac{1}{n}  \right)\\
&\times&  \exp\left[ +2\Delta^2(t-t_0)^2  \right]\ \frac{\gamma_b}{\Delta}\ \sqrt{\frac{\pi}{8}}\   e^{-\gamma\tau} \, . \nonumber
\end{eqnarray}
In the above calculation we assume that in the relevant time frame ($1/\gamma$) the first-order coherence is practically a constant ($G^{(1) }(0,t)\approx G^{(1) }(0,t+\tau)$), since $\gamma\gg \Delta$.
The result $g^{(2)}(0)$ diverges for $\Delta\rightarrow 0$ and for $t_0\rightarrow \infty$.
This is since here the detection of two photons occurs practically with the same probability as single photon. Lets mark this probability as $P$.
We then estimate $G^{(1)}(0)\propto P$ and  $G^{(2)}(0)\propto P$. This means $g^{(2)}(0)\propto P/P^2=1/P$. 
In the limits $\Delta\rightarrow 0$ and  $t_0\rightarrow \infty$ we have $P\rightarrow 0$ and therefore  $g^{(2)}(0)\rightarrow \infty$.
This is a well known result for bunched photons appearing with small probability.


\begin{thebibliography}{99}


\bibitem{Lvovsky2009}
A. I. Lvovsky and M. G. Raymer, Rev. Mod. Phys. {\bf 81}, 299 (2009).

\bibitem{WallsMilburn}
D. F. Walls and G. J. Milburn, {\em Quantum Optics} (Springer, Berlin, 2008).

\bibitem{Gisins}
N. Gisin, G. Ribordy, W. Tittel, and H. Zbinden, Rev. Mod. Phys. {\bf 74}, 145 (2002).


\bibitem{Menicucci2006}
N. C. Menicucci, P. van Loock, M. Gu, C. Weedbrook, T. C. Ralph, and M. A. Nielsen, Phys. Rev. Lett. {\bf 97}, 110501 (2006). 

\bibitem{Milburn2007}
P. Kok, W. J. Munro, K. Nemoto, T. C. Ralph, J. P. Dowling, and G. J. Milburn, Rev. Mod. Phys. {\bf 79}, 135 (2007).

\bibitem{Langford2011}
N. K. Langford, S. Ramelow, R. Prevedel, W. J. Munro, G. J. Milburn, and A. Zeilinger, Nature {\bf 478}, 360 (2011).

\bibitem{Solano2009}
G. Romero, J. J. Garcia-Ripoll, and E. Solano, Phys. Rev. Lett. {\bf 102}, 173602 (2009).


\bibitem{Chen2011}
Y.-F. Chen, D. Hover, S. Sendelbach, L. Maurer, S. T. Merkel, E. J. Pritchett, F. K. Wilhelm, and R. McDermott, Phys. Rev. Lett. {\bf 107}, 217401 (2011). 

\bibitem{Govia2012}
L. C. G. Govia, E. J. Pritchett, S. T. Merkel, D. Pineau, and F. K. Wilhelm, Phys. Rev. A {\bf 86}, 032311 (2012). 


\bibitem{Sathyamoorthy2014}
S. R. Sathyamoorthy, L. Tornberg, A. F. Kockum, B. Q. Baragiola, J. Combes, C. M. Wilson, T. M. Stace, and G. Johansson, Phys. Rev. Lett. {\bf 112}, 093601 (2014).

\bibitem{Koshino2015}
K. Koshino, K. Inomata, Z. Lin, Y. Nakamura, and T. Yamamoto, Phys. Rev. A {\bf 91}, 043805 (2015).

\bibitem{Inomata2016}
K. Inomata, Z. Lin, K. Koshino, W. D. Oliver, J.-S. Tsai, T. Yamamoto, Y. Nakamura, Nature Comm. {\bf 7}, 12303 (2016).

\bibitem{Kyrienko2016}
  O. Kyriienko and A. S. Sorensen, Phys. Rev. Lett. {\bf 117}, 140503 (2016).

\bibitem{Royer2017}
  B. Royer, A. L. Grimsmo, A. Choquette-Poitevin and A. Blais, arxiv:1710.06040v1 (2017).

\bibitem{Yurke1988}
  B. Yurke, P. G. Kaminsky, R. E. Miller, E. A. Whittaker, A. D. Smith, A. H. Silver and R. W. Simon, Phys. Rev. Let. {\bf 60}, 764 (1988).

\bibitem{Castellanos-Beltran2007}
  M. A. Castellanos-Beltran and K. W. Lehnert, Appl. Phys. Lett. {\bf 91}, 083509 (2007).

\bibitem{Bergeal2010}
  N. Bergeal, F. Schackert, M. Metcalfe, R. Vijay, V. E. Manucharyan, L. Frunzio, D. E. Prober, R. J. Schoelkopf, S. M. Girvin, M. H. Devoret, Nature, {\bf 465}, 64 (2010).

\bibitem{Macklin2015}
  C. Macklin, K. O'Brien,, D. Hover, M. E. Schwartz, V. Bolkhovsky, X. Zhang, W. D. Oliver, I. Siddiqi, Science {\bf 350}, 307 (2015).


\bibitem{Ingold1992}
G.-L.~Ingold and Yu.~V.~Nazarov, in {\em Single Charge Tunneling: Coulomb Blockade Phenomena in Nanostructures}, edited by H.~Grabert and M.~H.~Devoret (Plenum, New York, 1992), p.21.

\bibitem{Devoret1990}
M.~H.~Devoret, D.~Esteve, H.~Grabert, G.-L.~Ingold, H.~Pothier, and C.~Urbina, Phys.~Rev.~Lett.~{\bf 64}, 1824 (1990).

\bibitem{Girvin1990}
S. M. Girvin, L. I. Glazman, M. Jonson, D. R. Penn, and M. D. Stiles, Phys.~Rev.~Lett.~{\bf 64}, 3183 (1990).

\bibitem{Holst1994}
T.~Holst, D.~Esteve, C.~Urbina, and M.~H.~Devoret, Phys.~Rev.~Lett.~{\bf 73}, 3455 (1994).

\bibitem{Pertti2006}
J. Lepp\"akangas, E. Thuneberg, R. Lindell, and P. Hakonen, Phys. Rev. B {\bf 74} 054504 (2006).


  




\bibitem{Hofheinz2011}
M.~Hofheinz, F.~Portier, Q.~Baudouin, P.~Joyez, D.~Vion, P.~Bertet, P.~Roche, and D.~Esteve, Phys.~Rev.~Lett.~{\bf 106}, 217005 (2011).




\bibitem{Reulet3}
J.-C. Forgues, C. Lupien, and B. Reulet,  Phys. Rev. Lett. {\bf 114}, 130403 (2015).

\bibitem{Saira2016}
O.-P. Saira, M. Zgirski, K. L. Viisanen, D. S. Golubev, J. P. Pekola, Phys. Rev. Applied {\bf 6}, 024005 (2016).  


\bibitem{Parlavecchio2015}  
O. Parlavecchio, C. Altimiras, J.-R. Souquet, P. Simon, I. Safi, P. Joyez, D. Vion, P. Roche, D. Esteve, and F. Portier,
Phys. Rev. Lett.~{\bf 114} 126801 (2015).

\bibitem{Fabien2017}
M. Westig, B. Kubala, O. Parlavecchio, Y. Mukharsky, C. Altimiras, P Joyez, D. Vion, P. Roche, D. Esteve, M. Hofheinz, M. Trif, P. Simon, J. Ankerhold, and F. Portier, Phys. Rev. Lett. {\bf 119}, 137001 (2017).


\bibitem{Cassidy2017}
M. C. Cassidy, A. Bruno, S. Rubbert, M. Irfan, J. Kammhuber, R. N. Schouten, A. R. Akhmerov, and L. P.Kouwenhoven, Science {\bf 355}, 939 (2017).













\bibitem{Marthaler2011}
M. Marthaler, J. Lepp\"akangas, and J. H. Cole, Phys. Rev. B {\bf 83}, 180505(R) (2011).


\bibitem{Leppakangas2013}
J.~Lepp\"akangas, G.~Johansson, M.~Marthaler, and M. Fogelstr\"om, Phys.~Rev.~Lett.~{\bf 110}, 267004 (2013).



\bibitem{Gramich2013}
V.~Gramich, B.~Kubala, S.~Rohrer, J.~Ankerhold, Phys. Rev. Lett.~{\bf 111} 247002 (2013). 


\bibitem{Armour2013}
A.~D.~Armour, M.~P.~Blencowe, E.~Brahimi, A.~J.~Rimberg, Phys. Rev. Lett.~{\bf 111} 247001 (2013). 

\bibitem{Leppakangas2015}
J.~Lepp\"akangas, M. Fogelstr\"om, A.~Grimm, M.~Hofheinz, M.~Marthaler, and G.~Johansson, Phys. Rev. Lett.~{\bf 115} 027004 (2015).

\bibitem{Dambach2015}
S. Dambach, B. Kubala, V. Gramich, and J. Ankerhold,  Phys. Rev. B {\bf 92}, 054508 (2015).

\bibitem{Paris2015}
M. Trif and P. Simon, Phys. Rev. B {\bf 92}, 014503 (2015).


\bibitem{Quassemi2015}
A. L. Grimsmo, F. Qassemi, B. Reulet, and A. Blais, Phys. Rev. Lett. {\bf 116} 043602 (2016).

\bibitem{Hassler2015}
F.~Hassler and D.~Otten, Phys. Rev. B {\bf 92}, 195417 (2015).

\bibitem{Leppakangas2016}
J.~Lepp\"akangas, M. Fogelstr\"om, M.~Marthaler, and G.~Johansson, Phys. Rev. B {\bf 93} 014506 (2016).

\bibitem{Souquet2016}
J.-R. Souquet and A. A. Clerk, Phys. Rev. A {\bf 93}, 060301(R) (2016).

\bibitem{Koppenhofer2017}
M. Koppenh\"ofer, J. Lepp\"akangas, and M. Marthaler, Phys. Rev. B {\bf 95}, 134515 (2017).

\bibitem{DCE}
C. M. Wilson, G. Johansson, A. Pourkabirian, M. Simoen, J. R. Johansson, T. Duty, F. Nori, and P. Delsing, Nature {\bf 479}, 376 (2011).

\bibitem{Lahteenmaki2013}
P. L\"ahteenm\"aki, G. S. Paraoanu, J. Hassel, and P. J. Hakonen, Proc. Natl. Acad. Sci. U.S.A {\bf 110}, 4234 (2013).



\bibitem{WallquistPRB}
Wallquist~M, Shumeiko~V~S and Wendin~G 2006 {\it Phys.~Rev.}~B {\bf 74} 224506


\bibitem{Loudon}
R. Loudon, {\em The Quantum Theory of Light} (Oxford University, New York, 2010).

\bibitem{Fan2010}
S. Fan, S. E. Kocabas, and J.-T. Shen, Phys. Rev. A {\bf 82} 063821 (2010). 

\bibitem{Franson1989}
J. D. Franson, Phys. Rev. Lett. {\bf 62}, 2205 (1989).


\bibitem{Leppakangas2014}
J.~Lepp\"akangas, G.~Johansson, M.~Marthaler, and M. Fogelstr\"om, New J.~Phys.~{\bf 16}, 015015 (2014).








\bibitem{Wunsche1991}
A.\ W\"unsche, Quantum Opt.\ \textbf{3}, 359 (1991).






\bibitem{Eichler2011}
  C. Eichler, D. Bozyigit, C. Lang, L. Steffen, J. Fink, and A. Wallraff, Phys. Rev. Lett. {\bf 106}, 220503 (2011).

\bibitem{Kim97}
  M. S. Kim, Phys. Rev. A {\bf 56}, 3175 (1997).


\bibitem{Samkharadze2016}
  N. Samkharadze, A. Bruno, P. Scarlino, G. Zheng, D. P. DiVincenzo, L. DiCarlo and L. M. K. Vandersypen, Phys. Rev. Applied {\bf 5}, 044004 (2016).




\bibitem{FabienArray}
C. Altimiras, O. Parlavecchio, P. Joyez, D. Vion, P. Roche, and F. Portier, Appl. Phys. Lett {\bf 103}, 212601 (2013).


Appl. Phys. Lett. 103, 212601
(2013).

\bibitem{Stockklauser2017}
A. Stockklauser, P. Scarlino, J. V. Koski, S. Gasparinetti, C. K. Andersen, C. Reichl, W. Wegscheider, T. Ihn, K. Ensslin, and A. Wallraff,
Phys. Rev. X {\bf 7}, 011030 (2017).


\bibitem{Bundles2014}
C. S\'anchez Mu\~noz, E. del Valle, A. Gonz\'alez Tudela, K. M\"uller, S. Lichtmannecker, M. Kaniber, C. Tejedor, J. J. Finley, and F. P. Laussy,
Nature Photonics {\bf 8}, 550 (2014).


\bibitem{Anton2017}
A. Frisk Kockum, V. Macr\'i, L. Garziano, S. Savasta, and F. Nori, Scientific Reports {\bf 7}, 5313 (2017).

\bibitem{CatCodes} N. Ofek, A. Petrenko, R. Heeres, P. Reinhold, Z. Leghtas, B. Vlastakis, Y. Liu, L. Frunzio, S. M. Girvin, L. Jiang, M. Mirrahimi, M. H. Devoret, and R. J. Schoelkopf, Nature {\bf 536}, 441 (2016).

\end{thebibliography}
\end{document}